\documentclass[aps,onecolumn,letterpaper,oneside,preprint,tightenlines,draft,%
nobibnotes,nofootinbib,amsfonts,amssymb,amsmath,eqsecnum,showpacs,%
showkeys]{revtex4}

\newcommand{\cN}{\mathcal{N}}
\newcommand{\cV}{\mathcal{V}}
\newcommand{\cW}{\mathcal{W}}
\newcommand{\tF}{\tilde{F}}
\newcommand{\tH}{\tilde{H}}
 
\newcommand{\tJ}{\tilde{J}} 
\newcommand{\tK}{\tilde{K}} 
\newcommand{\tP}{\tilde{P}}

\newcommand{\tcV}{\tilde{\cV}}
\newcommand{\ta}{\tilde{a}}
\newcommand{\tb}{\tilde{b}}
\newcommand{\tw}{\tilde{w}}
\newcommand{\tvph}{\tilde{\varphi}}

\newcommand{\bbN}{\mathbb{N}}

\newcommand{\bbR}{\mathbb{R}}

\newcommand{\rme}{\mathrm{e}}
\newcommand{\rmd}{\mathrm{d}}
\newcommand{\rmA}{\mathrm{A}}
\newcommand{\rmB}{\mathrm{B}}
\newcommand{\rmC}{\mathrm{C}}

\newcommand{\braket}[1]{\left\langle{#1}\right\rangle}
\newcommand{\fsl}{\mathfrak{sl}}
\newcommand{\sqal}{\sqrt{\alpha}}
\newcommand{\sqab}{\sqrt{\alpha\beta}}

\begin{document}


%
%

\title{Type B 3-fold Supersymmetry and Non-polynomial Invariant Subspaces}
\author{Toshiaki Tanaka}
\email{tanaka.toshiaki@ocha.ac.jp}
\affiliation{Institute of Particle and Nuclear Studies,
 High Energy Accelerator Research Organization (KEK),
 1-1 Oho, Tsukuba, Ibaraki 305-0801, Japan}
\altaffiliation{Department of Physics,
 Faculty of Science, Ochanomizu University,
 2-1-1 Ohtsuka, Bunkyo-ku, Tokyo 112-8610, Japan}


\begin{abstract}

We obtain the most general type B 3-fold supersymmetry by solving directly
the intertwining relation. We then show that it is a necessary and sufficient
condition for a second-order linear differential operator to have three
linearly independent local analytic solutions. We find that there are eight 
linearly independent non-trivial linear differential operators of this kind.
As a by-product, we find new quasi-solvable second-order operators preserving
a monomial or polynomial subspace, one in type B, two in type C, and four in
type $X_{2}$, all of which have been missed in the existing literature.
In addition, we show that type A, type B, and type C 3-fold supersymmetries
are connected continuously via one parameter. A few new quasi-solvable
models are also presented.

\end{abstract}


\pacs{02.30.Hq; 03.65.Ca; 03.65.Ge; 11.30.Pb}
\keywords{$\cN$-fold supersymmetry; Quasi-solvability; Schr\"{o}dinger 
operators; Invariant spaces; Monomial subspaces; Polynomial subspaces}



\preprint{TH-1593}

\maketitle

\section{Introduction}
\label{sec:intro}

Since the discovery of the equivalence between \emph{$\cN$-fold supersymmetry}
(SUSY) and \emph{weak quasi-solvability}~\cite{AST01b} in 2001, the research
activity on exactly solvable Schr\"{o}dinger operators has come into
a new era. Indeed, we now recognize that all the exactly solvable ones
whose solutions are expressible in terms of a classical orthogonal polynomial
system are characterized as a particular case of type A $\cN$-fold
SUSY~\cite{ANST01,Ta03a} and that the ones which admit two linearly
independent series solutions belong to type C $\cN$-fold SUSY~\cite{GT05}.
In addition, a particular class of type B $\cN$-fold SUSY constructed in
Ref.~\cite{GT04} is now placed in a set of second-order linear
differential operators preserving a so-called exceptional polynomial
subspace of codimension one~\cite{GKM07,GKM10}. A set of new quasi-solvable
operators preserving an exceptional polynomial subspace of codimension
two was also constructed in the framework of $\cN$-fold SUSY~\cite{Ta10a},
which is called type $X_{2}$. For the development of $\cN$-fold SUSY until
the middle of 2000s and the terminology, see the review~\cite{Ta09}.
For its non-perturbative aspects, see also the review~\cite{Ta11b}.

The mathematical structure of $\cN$-fold SUSY is characterized by
a higher-order intertwining relation if it is represented by differential
operators. In general, however, it is quite difficult to solve directly
the relation under consideration which actually consists of simultaneous
coupled nonlinear differential equations. Until now on, the most general
form of $\cN$-fold SUSY is known only up to $\cN=4$~\cite{Ta11a}. They are
all weakly quasi-solvable, that is, they keep a linear subspace annihilated
by the corresponding $\cN$-fold supercharge invariant. It does not
immediately mean, however, that some local solutions are available in
a closed form. In other words, it does not automatically guarantee
quasi-solvability \emph{in the strong sense}.

Hence, a natural question would arise; under what conditions an $\cN$-fold
SUSY system admits a number of local analytic solutions. In the case where
the number is two, it was proved \cite{GT06} that a necessary and sufficient
condition for it is to have type A $2$-fold SUSY. The latter fact further
enabled one to construct systematically shape-invariant potentials within
this type of symmetry \cite{BT09,RT13}. These achievements clearly show
the power of $\cN$-fold SUSY, and we expect that we would be able to step
forward to more general cases, beginning with the case where the number of
available analytic solutions is three.

On the other hand, among the aforementioned four types of $\cN$-fold SUSY
which are different with each other for $\cN>2$, type A is the only one
for which a necessary and sufficient condition is known. It seems that
it is still quite hard to find it for an arbitrary $\cN>2$ even if we
restrict our consideration to a particular type . Hence, it would be
a realistic plan to make the examination from lower to higher values of
$\cN$ step by step until when we eventually discover, if it exists,
an inductive way toward a general statement applicable to an arbitrary
$\cN>2$ case.

In this paper, we examine under what conditions a one-body quantum
Hamiltonian would admit three linearly independent local solutions in
closed form. In addition, we investigate what type of $3$-fold SUSY
such a system must possess as a consequence of the equivalence between
$\cN$-fold SUSY and weak quasi-solvability. We show that it is the most
general type B, for which only a particular case has been so far explored.
In other words, we prove that type B $3$-fold SUSY is a necessary and
sufficient condition for a Hamiltonian to admit three linearly independent
local analytic solutions. We find that there are eight linearly independent
second-order linear differential operators of this kind. As a by-product
of this finding, we obtain some new quasi-solvable operators preserving
a three-dimensional monomial or polynomial subspace which have not been
considered in the literature. More precisely, we discover one new
quasi-solvable operator for a type B monomial space, two for type C,
and four for type $X_{2}$.
Furthermore, we find that all the eight quasi-solvable operators in
the type A, type B, and type C cases are connected continuously via
a one parameter.

We organize the paper as follows. In Section~\ref{sec:qs3d}, we first
construct the most general second-order linear differential operator which
preserves a three-dimensional linear function space. In Section~\ref{sec:B3},
we solve directly the intertwining relation in type B $3$-fold SUSY to obtain
a necessary and sufficient condition for it. We then show that it leads to
the operator which is identical with the one obtained in Section~\ref{sec:qs3d}. 
Sections~\ref{sec:Ex} and \ref{sec:ABC} deal with particular cases of the
most general type B SUSY. In Section~\ref{sec:Ex}, we present three
examples of type B 3-fold SUSY models obtained by a particular choice of
the three-dimensional linear function space. In Section~\ref{sec:ABC},
we consider the cases where the three-dimensional linear space is given
by type A, type B, and type C monomial subspaces. We show there that
these three different types are continuously connected via a parameter.
We also find new quasi-solvable operators preserving type B and type C
monomial subspaces. In Section~\ref{sec:CR},
we examine commutation relations of quasi-solvable operators.
The analysis demonstrates a reason why the family of type A
models is special in the sense that it is Lie-algebraic.
In Section~\ref{sec:X2}, we first reformulate the most general type B
$3$-fold SUSY system in Section~\ref{sec:B3}, and then apply it to
the case of three-dimensional $X_{2}$ polynomial subspaces.
In Section~\ref{sec:discus}, we summarize the results and discuss their
implications and prospects for the future studies.

\section{Quasi-solvable Operators preserving a Three-dimensional Space}
\label{sec:qs3d}

Let us first consider a one-dimensional quantum Hamiltonian $H$ which admits
three linearly independent local analytic solutions in closed form to the 
corresponding Schr\"{o}dinger equation,
\begin{align}
H\psi_{i}(q)=E_{i}\psi_{i}(q)\qquad (i=1,2,3),
\label{eq:Sch}
\end{align}
where each of $\psi_{i}$ does not necessarily normalizable. It is evident
that the Hamiltonian $H$ preserves a three-dimensional linear space $\cV_{3}$
of functions defined by
\begin{align}
\cV_{3}=\braket{\psi_{1}(q),\psi_{2}(q),\psi_{3}(q)}.
\label{eq:V3}
\end{align}
Conversely, if a Hamiltonian preserves a linear space (\ref{eq:V3}) spanned
by three known functions, we can diagonalize it in the three-dimensional space 
algebraically, and thus obtain three linearly independent local solutions to
the corresponding Schr\"{o}dinger equation. An available set of three functions
differs depending on
the Hamiltonian under consideration, but we can extract a common mathematical
structure shared by all such systems. For this purpose, let us introduce a 
``gauged'' Hamiltonian $\tH^{-}$ by $\tH^{-}=\psi_{1}(q)H\psi_{1}(q)^{-1}$.
Then, it immediately follows from the invariance $H\cV_{3}\subset\cV_{3}$
that the gauged Hamiltonian $\tH^{-}$ preserves the linear space
\begin{align}
\tcV_{3}^{-}[z]=\braket{1,z,f(z)},
\label{eq:tV3-}
\end{align}
where the variable $z$ and the function $f(z)$ are defined by
\begin{align}
z=\psi_{2}(q)/\psi_{1}(q),\qquad f(z)=\psi_{3}(q)/\psi_{1}(q).
\end{align}
It is evident from the assumed linear independence of $\psi_{i}(q)$
($i=1,2,3$) that $f(z)$ is arbitrary with the exception of a polynomial
of at most first degree in $z$. That is, $f(z)$ is an arbitrary function
satisfying $f''(z)\neq 0$. In addition, it is not necessary for
the function $f(z)$ to admit an explicit form in terms of the variable
$z$. As we shall see in Section~\ref{sec:B3}, the specific choice of the
variable $z$, which results in the form of the vector space (\ref{eq:tV3-}),
enables us to establish easily a connection with an existing type of 3-fold
SUSY. Later in Section~\ref{sec:X2}, we shall reconsider the problem with
the most general change of variable.

{}From the above argument, it is now clear that any one-dimensional
quantum Hamiltonian $H$ in the physical $q$-space which has three
linearly independent local analytic solutions can be converted with
a gauge transformation and a change of variable $z=z(q)$ to a gauged
Hamiltonian $\tH^{-}$ in the gauged $z$-space which leaves the linear 
space (\ref{eq:tV3-}) invariant. Such a gauged Hamiltonian $\tH^{-}$
is no longer a Schr\"{o}dinger operator but is a general linear
second-order differential operator in the variable $z$:
\begin{align}
\tH^{-}=-A(z)\frac{\rmd^{2}}{\rmd z^{2}}-B(z)\frac{\rmd}{\rmd z}-C(z),
\label{eq:tH-}
\end{align}
where $A(z)$, $B(z)$, and $C(z)$ are functions of $z$ to be determined. To 
preserve the space (\ref{eq:tV3-}), it is necessary and sufficient for
$\tH^{-}$ to satisfy
\begin{align}
\tH^{-}1&=-C(z)=-c_{2}f(z)-c_{1}z-c_{0},\\
\tH^{-}z&=-B(z)-C(z)z=-b_{2}f(z)-b_{1}z-b_{0},\\
\tH^{-}f(z)&=-A(z)f''(z)-B(z)f'(z)-C(z)f(z)=-a_{2}f(z)-a_{1}z-a_{0},
\end{align}
where $c_{i}$, $b_{i}$, and $a_{i}$ ($i=0,1,2$) are all constants. From these 
conditions, we obtain
\begin{align}
A(z)f''(z)=&\;\left[(c_{2}z-b_{2})f(z)+c_{1}z^{2}+(c_{0}-b_{1})z-b_{0}
 \right]f'(z)\notag\\
&\;-[c_{2}f(z)+c_{1}z+c_{0}-a_{2}]f(z)+a_{1}z+a_{0},
\label{eq:Az}\\
B(z)=&\;-(c_{2}z-b_{2})f(z)-c_{1}z^{2}-(c_{0}-b_{1})z+b_{0},\\
C(z)=&\;c_{2}f(z)+c_{1}z+c_{0}.
\label{eq:Cz}
\end{align}
Substituting the latter formulas for $A(z)$, $B(z)$, and $C(z)$ into
(\ref{eq:tH-}), we see that $\tH^{-}$ is expressed as
\begin{align}
\tH^{-}=&\;-c_{2}J_{8}-c_{1}J_{7}+b_{2}J_{6}+(b_{1}-c_{0})J_{5}
 +b_{0}J_{4}\notag\\
&\;-(a_{2}-c_{0})J_{3}-a_{1}J_{2}-a_{0}J_{1}-c_{0},
\label{eq:tH-2}
\end{align}
where $J_{i}$ ($i=1,\dots,8$) are given by
\begin{align}
\begin{split}
&J_{1}[z]=\frac{1}{f''(z)}\frac{\rmd^{2}}{\rmd z^{2}},\quad J_{2}[z]=
 z J_{1}[z]=\frac{z}{f''(z)}\frac{\rmd^{2}}{\rmd z^{2}},\quad J_{3}[z]=
 f(z)J_{1}[z]=\frac{f(z)}{f''(z)}\frac{\rmd^{2}}{\rmd z^{2}},\\
&J_{4}[z]=\frac{f'(z)}{f''(z)}\frac{\rmd^{2}}{\rmd z^{2}}-\frac{\rmd}{\rmd z},
 \qquad J_{5}[z]=z J_{4}[z]=\frac{z f'(z)}{f''(z)}\frac{\rmd^{2}}{\rmd z^{2}}
 -z\frac{\rmd}{\rmd z},\\
&J_{6}[z]=f(z)J_{4}[z]=\frac{f(z)f'(z)}{f''(z)}\frac{\rmd^{2}}{\rmd z^{2}}
 -f(z)\frac{\rmd}{\rmd z},\\
&J_{7}[z]=z J_{9}[z]=\frac{z f'(z)-f(z)}{f''(z)}z\frac{\rmd^{2}}{\rmd z^{2}}
 -z^{2}\frac{\rmd}{\rmd z}+z,\\
&J_{8}[z]=f(z)J_{9}[z]=\frac{z f'(z)-f(z)}{f''(z)}f(z)\frac{\rmd^{2}}{
 \rmd z^{2}}-z f(z)\frac{\rmd}{\rmd z}+f(z),
\end{split}
\label{eq:Js}
\end{align}
and $J_{9}$ is defined by
\begin{align}
J_{9}[z]=J_{5}[z]-J_{3}[z]+1=\frac{z f'(z)-f(z)}{f''(z)}
 \frac{\rmd^{2}}{\rmd z^{2}}-z\frac{\rmd}{\rmd z}+1.
\end{align}
Therefore, a linear space of non-trivial differential operators of at most 
second order which preserves the space (\ref{eq:tV3-}) is spanned by the
eight linearly independent operators $J_{i}$ ($i=1,\dots,8$). This result
is indeed consistent with the general ones studied in Ref.~\cite{KMO00}.

It is worth noting that there are in general no first-order linear
differential operators which preserves (\ref{eq:tV3-}). In fact, we
can easily see that Eq.~(\ref{eq:Az}) provides a constraint
on the function $f(z)$ when we set $A(z)=0$ to consider a first-order
quasi-solvable operator. This constraint cannot be satisfied unless
$f(z)$ has certain specific forms, cf.\ Sections~\ref{sec:ABC}
and~\ref{sec:CR}.

\section{Type B 3-fold SUSY}
\label{sec:B3}

Type B $\cN$-fold SUSY was first discovered in Ref.~\cite{GT04} by a simple
deformation of type A $\cN$-fold supercharge.
The component of type B $3$-fold supercharge is given by
\begin{align}
P_{3}^{-}=\left(\frac{\rmd}{\rmd q}+W(q)-E(q)-F(q)\right)\left(
 \frac{\rmd}{\rmd q}+W(q)\right)\left(\frac{\rmd}{\rmd q}+W(q)+E(q)\right),
\label{eq:P3-}
\end{align}
where the three functions $W(q)$, $E(q)$, and $F(q)$ are at present arbitrary.
A pair of Hamiltonians,
\begin{align}
H^{\pm}=-\frac{1}{2}\frac{\rmd^{2}}{\rmd q^{2}}+V^{\pm}(q),
\label{eq:H+-}
\end{align}
is said to have type B $3$-fold SUSY if it is intertwined by $P_{3}^{-}$
in (\ref{eq:P3-}) as
\begin{align}
P_{3}^{\mp}H^{\mp}=H^{\pm}P_{3}^{\mp},
\label{eq:inter}
\end{align}
where $P_{3}^{+}$ is the transposition of $P_{3}^{-}$ in the $q$-space, 
$P_{3}^{+}=(P_{3}^{-})^{\mathrm{T}}$. A direct calculation shows that the 
intertwining relation (\ref{eq:inter}) holds if and only if the potential
terms in (\ref{eq:H+-}) have the following form
\begin{align}
V^{\pm}=\frac{1}{2}W^{2}-\frac{1}{3}\left( 2 E'-E^{2}\right)-\frac{1}{6}
 \left( 2 F'+2 W\! F-2 E F-F^{2}\right)\pm\frac{1}{2}\left(3 W'-F'\right),
\label{eq:cond1}
\end{align}
and simultaneously the three functions $W(q)$, $E(q)$, and $F(q)$ satisfy
\begin{align}
\left(\frac{\rmd}{\rmd q}-E\right) F'_{1}-\frac{F}{2}\left( F'_{1}
 -\frac{F'_{2}}{6}\right) =0,
\label{eq:cond2}\\
\left(\frac{\rmd}{\rmd q}-2 E-\frac{3}{2}F\right)\left(\frac{\rmd}{\rmd q}-E
 \right) F'_{2}+\frac{3}{2}\left( 2 F'-2 E F-F^{2}\right)\left( F'_{1}
 -\frac{F'_{2}}{6}\right) =0,
\label{eq:cond3}
\end{align}
where $F_{1}(q)$ and $F_{2}(q)$ are given by
\begin{align}
\begin{split}
F_{1}&=W'+E W-\frac{1}{4}\left( F'-2 W\! F+2 E F+F^{2}\right),\\
F_{2}&=E'+E^{2}+\frac{1}{2}\left( F'-2 W\! F+2 E F+F^{2}\right).
\end{split}
\label{eq:F1F2}
\end{align}
We note that the factorized form of the type B $3$-fold supercharge component
(\ref{eq:P3-}) is expanded as
\begin{align}
P_{3}^{-}=\frac{\rmd^{3}}{\rmd q^{3}}+\sum_{k=0}^{2}w_{k}^{[3]}(q)
 \frac{\rmd^{k}}{\rmd q^{k}},
\label{eq:gP3-}
\end{align}
with
\begin{subequations}
\label{eqs:wk3}
\begin{align}
w_{2}^{[3]}=&\;3 W-F,\\
w_{1}^{[3]}=&\;3 W'+2 E'+3 W^{2}-E^{2}-2 W\! F-E F,\\
w_{0}^{[3]}=&\;W''+E''+3 WW'+2 E'W-E E'-W'F-E'F\notag\\
&\;+W^{3}-E^{2}W-W^{2}F-E W\! F.
\end{align}
\end{subequations}
Then, the same conditions (\ref{eq:cond1})--(\ref{eq:F1F2}) can be also
derived from the general form of $3$-fold SUSY in Ref.~\cite{Ta11a} by 
substituting (\ref{eqs:wk3}) into the formulas in the latter reference.

To solve the coupled differential equations (\ref{eq:cond2}) and
(\ref{eq:cond3}), it is convenient to make a change of variable $z=z(q)$
and to introduce a function $f(z)$ defined by
\begin{align}
E(q)=\frac{z''(q)}{z'(q)},\qquad F(q)=\frac{f'''(z(q))}{f''(z(q))}z'(q).
\label{eq:EF}
\end{align}
In terms of the new variable $z$ and the function $f(z)$, Eqs.~(\ref{eq:cond2})
and (\ref{eq:cond3}) are converted into
\begin{align}
\tF''_{1}(z)-\frac{f'''(z)}{2 f''(z)}\left( \tF'_{1}(z)-\frac{\tF'_{2}(z)}{6}
 \right) =0,
\label{eq:cond2'}\\
\left(\frac{\rmd}{\rmd z}-\frac{3 f'''(z)}{2 f''(z)}\right) \tF''_{2}(z)
 +\frac{3}{2}\left( \frac{2 f''''(z)}{f''(z)}-\frac{3 f'''(z)^{2}}{f''(z)^{2}}
 \right) \left( \tF'_{1}(z)-\frac{\tF'_{2}(z)}{6}\right)=0,
\label{eq:cond3'}
\end{align}
where $\tF_{i}(z(q))=F_{i}(q)$ ($i=1,2$). Eliminating $\tF_{2}(z)$ from
(\ref{eq:cond2'}) and (\ref{eq:cond3'}), we obtain
\begin{align}
\frac{\rmd}{\rmd z}\left(\frac{f''(z)}{f'''(z)}\tF'''_{1}(z)-\frac{
 f''(z)f''''(z)}{f'''(z)^{2}}\tF''_{1}(z)\right) -\left( \tF'''_{1}(z)
 -\frac{f''''(z)}{f'''(z)}\tF''_{1}(z)\right) =0.
\end{align}
This differential equation for $\tF_{1}(z)$ can be integrated four times
to yield
\begin{align}
\tF_{1}(z)=C_{1}\left( z f'(z)-2 f(z)\right) +C_{2}f'(z)+C_{3}z+C_{4},
\label{eq:tF1}
\end{align}
where $C_{i}$ ($i=1,\dots,4$) are integral constants. Substituting it into
(\ref{eq:cond2'}) and integrating the resultant differential equation for 
$\tF_{2}(z)$, we have
\begin{align}
\tF_{2}(z)=-6(C_{1}z f'(z)+C_{2}f'(z)-C_{3}z-C_{5}),
\label{eq:tF2}
\end{align}
where $C_{5}$ is another integral constant. From the definition (\ref{eq:EF}),
we can express $z$ and $f(z(q))$ in terms of $E(q)$ and $F(q)$ as
\begin{align}
z(q)=\int\rmd q\,\rme^{\,\int^{q}\rmd q' E(q')},\quad f(z(q))=\int\rmd q
 \left[\rme^{\,\int^{q}\rmd q' E(q')}\int^{q}\rmd q' \rme^{\,\int^{q'}\rmd 
 q''(E(q'')+F(q''))}\right].
\end{align}
Substituting them into (\ref{eq:tF1}) and (\ref{eq:tF2}), we obtain
expressions for $F_{1}(q)$ and $F_{2}(q)$ defined in (\ref{eq:F1F2}).
Finally, substituting the resulting $F_{1}(q)$ and $F_{2}(q)$ into
(\ref{eq:F1F2}), we obtain the entire relations to be satisfied among
the three functions $W(q)$, $E(q)$, and $F(q)$. For instance, the coupled
differential equations (\ref{eq:F1F2}) can be integrated for $W(q)$ as
\begin{align}
W(q)=\rme^{-\int\rmd q\,E(q)}\int\rmd q\,\rme^{\,\int^{q}\rmd q'E(q')}\left(
 F_{1}(q)+\frac{F_{2}(q)-E'(q)-E(q)^{2}}{2}\right).
\label{eq:W-EF}
\end{align}
Thus, we can express $W(q)$ in terms of $E(q)$ and $F(q)$ by the substitution
for the obtained $F_{1}(q)$ and $F_{2}(q)$ into (\ref{eq:W-EF}). Then, 
eliminating $W(q)$ in (\ref{eq:F1F2}) by using (\ref{eq:W-EF}), we have
a relation between $E(q)$ and $F(q)$. In this way, we can obtain the most 
general form of type B $3$-fold SUSY.

Next, we shall show that type B $3$-fold SUSY systems are essentially 
identical with second-order linear differential operators preserving the
space (\ref{eq:tV3-}) investigated in Section~\ref{sec:qs3d}. For this
purpose, let us first make a ``gauge'' transformation on $H^{-}$:
\begin{align}
\tH^{-}=\rme^{\cW_{3}^{-}}H^{-}\rme^{-\cW_{3}^{-}},
\label{eq:gtH-}
\end{align}
with the gauge potential $\cW_{3}^{-}(q)$ given by
\begin{align}
\cW_{3}^{\pm}(q)=\int\rmd q\,( E(q)\mp W(q)).
\label{eq:cW3+-}
\end{align}
Substituting (\ref{eq:cond1}) for $V^{-}(q)$ into (\ref{eq:H+-}), we have
\begin{align}
\tH^{-}=-\frac{1}{2}\frac{\rmd^{2}}{\rmd q^{2}}+(W(q)+E(q))\frac{\rmd}{\rmd q}
 -F_{1}(q)-\frac{F_{2}(q)}{6},
\end{align}
where $F_{1}(q)$ and $F_{2}(q)$ are given by (\ref{eq:F1F2}). Thus, if we
make a change of variable $z=z(q)$ and introduce two functions $A(z)$ and
$Q(z)$ as
\begin{align}
2 A(z(q))=z'(q)^{2},\qquad Q(z(q))=-z'(q)W(q),
\label{eq:AQ}
\end{align}
we can express the gauged Hamiltonian $\tH^{-}$ in terms of $z$ as
\begin{align}
\tH^{-}=-A(z)\frac{\rmd^{2}}{\rmd z^{2}}-\left( Q(z)-\frac{A'(z)}{2}\right)
 \frac{\rmd}{\rmd z}-\tF_{1}(z)-\frac{\tF_{2}(z)}{6}.
\label{eq:tH-B}
\end{align}
Comparing it with (\ref{eq:tH-}), we immediately see the correspondence
between them as
\begin{align}
B(z)=Q(z)-\frac{A'(z)}{2},\qquad C(z)=\tF_{1}(z)+\frac{\tF_{2}(z)}{6}.
\label{eq:BC}
\end{align}
On the other hand, we can convert the formulas among the functions of $q$
in (\ref{eq:F1F2}) into the ones among the functions of $z$ by
employing (\ref{eq:EF}) and (\ref{eq:AQ}) as
\begin{align}
\begin{split}
\tF_{1}(z)&=-Q'(z)-\frac{2 f''''(z)A(z)+f'''(z)(3 A'(z)+2 Q(z))}{4 f''(z)},\\
\tF_{2}(z)&=A''(z)+\frac{2 f''''(z)A(z)+f'''(z)(3 A'(z)+2 Q(z))}{2 f''(z)}.
\end{split}
\label{eq:tF1F2}
\end{align}
Substituting (\ref{eq:tF1F2}) into (\ref{eq:tF1}) and (\ref{eq:tF2}), we
have coupled differential equations for $A(z)$ and $Q(z)$. The latter
equations can be easily integrated out to yield
\begin{align}
2 B(z)=&\;2 Q(z)-A'(z)\notag\\
=&\;4 C_{1}z f(z)+4 C_{2}f(z)-4 C_{3}z^{2}-2(C_{4}+3 C_{5})z-2 C_{6},
\label{eq:BzB}\\
A(z)f''(z)=&\;f'(z)\left[ -2 C_{1}z f(z)-2 C_{2}f(z)+2 C_{3}z^{2}+(C_{4}
 +3 C_{5})z+C_{6}\right]\notag\\
&\;+2 f(z)(C_{1}f(z)-C_{3}z-C_{4})+C_{7}z+C_{8},
\end{align}
where $C_{i}$ ($i=6,7,8$) are additional integral constants. The explicit
form of $C(z)$ can be obtained by substituting (\ref{eq:tF1}) and
(\ref{eq:tF2}) into the second formula in (\ref{eq:BC}) as
\begin{align}
C(z)=-2 C_{1}f(z)+2 C_{3}z+C_{4}+C_{5}.
\label{eq:CzB}
\end{align}
Comparing (\ref{eq:BzB})--(\ref{eq:CzB}) with (\ref{eq:Az})--(\ref{eq:Cz}),
we easily see that they are identical with each other and the parameters
are related by
\begin{align}
\begin{split}
&c_{2}=-2 C_{1},\qquad c_{1}=2 C_{3},\qquad c_{0}=C_{4}+C_{5},\qquad
 b_{2}=2 C_{2},\\
&b_{1}=-2 C_{5},\quad b_{0}=-C_{6},\quad a_{2}=C_{5}-C_{4},\quad
 a_{1}=C_{7},\quad a_{0}=C_{8}.
\end{split}
\label{eq:param}
\end{align}
Therefore, the gauged type B $3$-fold SUSY Hamiltonian (\ref{eq:tH-B})
exactly coincides with the most general second-order linear differential 
operator (\ref{eq:tH-2}), which leaves the linear space (\ref{eq:tV3-})
invariant. The latter fact can be also derived directly by using the type
B $3$-fold supercharge without recourse to the explicit form of $\tH^{-}$.
In the gauged $z$-space, the component of type B $3$-fold supercharge
reads as
\begin{align}
\tP_{3}^{-}[z]=\rme^{\cW_{3}^{-}}P_{3}^{-}\rme^{-\cW_{3}^{-}}
 =z'(q)^{3}\left(\frac{\rmd}{\rmd z}-\frac{f'''(z)}{f''(z)}\right)
 \frac{\rmd^{2}}{\rmd z^{2}},
\label{eq:tP3-}
\end{align}
and it actually annihilates all the elements of the linear space
(\ref{eq:tV3-}). In other words,
\begin{align}
\tcV_{3}^{-}=\ker \tP_{3}^{-}.
\label{eq:kertP3-}
\end{align}
On the other hand, it follows from the gauge-transformed version of the
intertwining relation (\ref{eq:inter}) with upper signs, namely, 
$\tP_{3}^{-}\tH^{-}=\tH^{+}\tP_{3}^{-}$, that $\tH^{-}\ker \tP_{3}^{-}
\subset\ker \tP_{3}^{-}$. Hence, the equality (\ref{eq:kertP3-})
indeed means that any gauged type B $3$-fold SUSY Hamiltonian $\tH^{-}$
defined by (\ref{eq:gtH-}) with (\ref{eq:cW3+-}) preserves the linear
space (\ref{eq:tV3-}).

One of the most advantageous and powerful aspects of the framework of
$\cN$-fold SUSY is that we can construct simultaneously another weakly 
quasi-solvable Hamiltonian $H^{+}$ which is almost isospectral with $H^{-}$. 
According to Ref.~\cite{GT05}, the partner Hamiltonian $H^{+}$ of $3$-fold
SUSY in the physical $q$-space is connected to the gauged $z$-space by
another gauge transformation induced by the gauge potential $\cW_{3}^{+}(q)$
already defined in (\ref{eq:cW3+-}). With the latter gauge transformation, 
$H^{+}$ is transformed to a second-order linear differential operator
$\bar{H}^{+}$ of the variable $z$ which must have the following expression:
\begin{align}
\bar{H}^{+}=&\;\rme^{\cW_{3}^{+}}H^{+}\rme^{-\cW_{3}^{+}}\notag\\
=&\;-A(z)\frac{\rmd^{2}}{\rmd z^{2}}+\left( \frac{A'(z)}{2}+Q(z)\right)
 \frac{\rmd}{\rmd z}-C(z)-2 Q'(z)\notag\\
&\;+A'(z)\tw_{2}^{[3]}(z)+2 A(z)\tw_{2}^{[3]\prime}(z),
\label{eq:bH+}
\end{align}
where $A(z)$ and $Q(z)$ are the same functions as the ones in (\ref{eq:AQ}), 
$C(z)$ is the zeroth-order term of $\tH^{-}$ in (\ref{eq:tH-}) and thus is 
given by the second formula in (\ref{eq:BC}) in our type B case, and 
$\tw_{2}^{[3]}(z)$ is defined via the expanded form of the gauged $3$-fold 
supercharge component $\tP_{3}^{-}$ under consideration:
\begin{align}
\tP_{3}^{-}=z'(q)^{3}\left(\frac{\rmd^{3}}{\rmd z^{3}}+\sum_{k=0}^{2}
 \tw_{k}^{[3]}(z)\frac{\rmd^{k}}{\rmd z^{k}}\right).
\end{align}
Thus, in our type B case (\ref{eq:tP3-}), it is given by
\begin{align}
\tw_{2}^{[3]}(z)=-\frac{f'''(z)}{f''(z)}.
\label{eq:tw2}
\end{align}
Substituting (\ref{eq:BzB})--(\ref{eq:param}) and (\ref{eq:tw2}) into
(\ref{eq:bH+}), we can express $\bar{H}^{+}$ as
\begin{align}
\bar{H}^{+}=&\;-c_{2}K_{8}-c_{1}K_{7}+b_{2}K_{6}+(b_{1}-c_{0})K_{5}+b_{0}
 K_{4}\notag\\
&\;-(a_{2}-c_{0})K_{3}-a_{1}K_{2}-a_{0}K_{1}-c_{0},
\label{eq:bH+2}
\end{align}
where the operators $K_{i}$ ($i=1,\dots,8$) are given by
\begin{align}
\begin{split}
&K_{1}[z]=\frac{1}{f''(z)}\frac{\rmd^{2}}{\rmd z^{2}}+\frac{f'''(z)}{
 f''(z)^{2}}\frac{\rmd}{\rmd z}+\frac{f''(z)f''''(z)-f'''(z)^{2}}{
 f''(z)^{3}},\\
&K_{2}[z]=z K_{1}[z]-K_{0}[z],\qquad K_{3}[z]=f(z)K_{1}[z]-f'(z)
 K_{0}[z]+1,\\[8pt]
&K_{4}[z]=f'(z)K_{1}[z],\qquad K_{5}[z]=f'(z)K_{2}[z],\qquad
 K_{6}[z]=f'(z)K_{3}[z],\\[8pt]
&K_{7}[z]=(z f'(z)-f(z))K_{2}[z],\qquad K_{8}[z]=(z f'(z)-f(z))K_{3}[z],
\end{split}
\label{eq:Ks}
\end{align}
and $K_{0}$ is defined by
\begin{align}
K_{0}[z]=\frac{1}{f''(z)}\frac{\rmd}{\rmd z}+\frac{f'''(z)}{f''(z)^{2}}.
\end{align}
By applying the formulas (\ref{eq:F1F2}), (\ref{eq:EF}), (\ref{eq:AQ}),
(\ref{eq:BC}), and (\ref{eq:tw2}), the gauged partner Hamiltonian
$\bar{H}^{+}$ in (\ref{eq:bH+}) is expressed in terms of $q$ as
\begin{align}
\bar{H}^{+}=&\;-\frac{1}{2}\frac{\rmd^{2}}{\rmd q^{2}}+(E-W)
 \frac{\rmd}{\rmd q}+W'+E W\notag\\
&\;-\frac{1}{6}\left( E'+E^{2}+5 F'+2 W\! F-2 E F-F^{2}\right).
\end{align}
Then, we can easily check that the potential term in the original Hamiltonian
$H^{+}=\rme^{-\cW_{3}^{+}}\bar{H}^{+}\rme^{\cW_{3}^{+}}$ coincides exactly with
$V^{+}(q)$ in (\ref{eq:cond1}), which was derived from the direct calculation 
of the intertwining relation (\ref{eq:inter}).

The solvable sector $\bar{\cV}_{3}^{+}$ preserved by $\bar{H}^{+}$ is 
characterized by the gauge-transformed operator of the transposed component 
$P_{3}^{+}$ of type B $3$-fold supercharge:
\begin{align}
\bar{P}_{3}^{+}=\rme^{\cW_{3}^{+}}P_{3}^{+}\rme^{-\cW_{3}^{+}}
 =-z'(q)^{3}\frac{\rmd^{2}}{\rmd z^{2}}\left(\frac{\rmd}{\rmd z}+\frac{
 f'''(z)}{f''(z)}\right).
\label{eq:bP3+}
\end{align}
In fact, the gauge transformation with $\cW_{3}^{+}$ of the intertwining
relation (\ref{eq:inter}) with lower signs, namely, 
$\bar{P}_{3}^{+}\bar{H}^{+}=\bar{H}^{-}\bar{P}_{3}^{+}$, immediately
leads to $\bar{H}^{+}\ker \bar{P}_{3}^{+}\subset\ker\bar{P}_{3}^{+}$.
Hence, we obtain from (\ref{eq:bP3+}):
\begin{align}
\bar{\cV}_{3}^{+}[z]=\ker \bar{P}_{3}^{+}=\frac{1}{f''(z)}
 \braket{1,f'(z),z f'(z)-f(z)}.
\label{eq:bV3+}
\end{align}
We can easily check that the eight linearly independent operators $K_{i}$
($i=1,\dots,8$) in (\ref{eq:Ks}) indeed preserve the linear space
(\ref{eq:bV3+}). It is worth noting that the space $\bar{\cV}_{3}^{+}$ in
(\ref{eq:bV3+}) is related to the one $\tcV_{3}^{-}$ in (\ref{eq:tV3-}) by
\begin{align}
\bar{\cV}_{3}^{+}[z]=f''(z)^{-1}\tcV_{3}^{-}[w]\bigr|_{w=f'(z),\, f(w)=z 
 f'(z)-f(z)}\,.
\end{align}
Then, it follows from this relation that a set of eight linearly independent
second-order linear differential operators preserving $\bar{\cV}_{3}^{+}$
can be constructed directly from $J_{i}$ ($i=1,\dots,8$) in (\ref{eq:Js}) as
\begin{align}
f''(z)^{-1}J_{i}[w]f''(z)\bigr|_{w=f'(z),\, f(w)=z f'(z)-f(z)}\quad
 (i=1,\dots,8).
\label{eq:Jws}
\end{align}
Actually, we can check that all the operators $K_{i}$ ($i=1,\dots,8$)
in (\ref{eq:Ks}) are expressed as linear combinations of the ones in
(\ref{eq:Jws}) as
\begin{align}
\begin{split}
K_{1}[z]&=f''(z)^{-1}J_{1}[w]f''(z),\qquad
 K_{2}[z]=f''(z)^{-1}J_{4}[w]f''(z),\\
K_{3}[z]&=f''(z)^{-1}(J_{5}[w]-J_{3}[w]+1)f''(z),\quad  
 K_{4}[z]=f''(z)^{-1}J_{2}[w]f''(z),\\
K_{5}[z]&=f''(z)^{-1}J_{5}[w]f''(z),\qquad
 K_{6}[z]=f''(z)^{-1}J_{7}[w]f''(z),\\
K_{7}[z]&=f''(z)^{-1}J_{6}[w]f''(z),\qquad
 K_{8}[z]=f''(z)^{-1}J_{8}[w]f''(z),
\end{split}
\label{eq:JKs}
\end{align}
where the substitution $w=f'(z)$ and $f(w)=z f'(z)-f(z)$ in each operator 
$J_{i}[w]$ has been understood.

By construction, it is evident that each of the pair Hamiltonians $H^{\pm}$,
which is connected with the gauged ones $\tH^{-}$ and $\bar{H}^{+}$ via
(\ref{eq:gtH-}) and (\ref{eq:bH+}), respectively, preserves a
three-dimensional vector space $\cV_{3}^{\pm}[q]$ defined by
\begin{align}
\begin{split}
\cV_{3}^{-}[q]&=\tcV_{3}^{-}[z(q)]\,\rme^{-\cW_{3}^{-}(q)},\\
\cV_{3}^{+}[q]&=\bar{\cV}_{3}^{+}[z(q)]\,\rme^{-\cW_{3}^{+}(q)},
\end{split}
\label{eq:cV3}
\end{align}
where $\tcV_{3}^{-}[z]$ and $\bar{\cV}_{3}^{+}[z]$ is respectively given
by (\ref{eq:tV3-}) and (\ref{eq:bV3+}). The linear space $\cV_{3}^{\pm}[q]$
is called a \emph{solvable sector} of $H^{\pm}$. If it is in addition,
a subspace of the linear space in which the corresponding Hamiltonian
acts, e.g., $L^{2}(\bbR)$ in a usual quantum mechanical setting, it is
\emph{quasi-exactly solvable} in the space.

\section{Examples of Type B 3-fold SUSY Models}
\label{sec:Ex}

To demonstrate what kind of quantum Hamiltonians can possess type B
3-fold SUSY, we shall exhibit some examples by choosing a particular
function for $f(z)$.

Let us consider type B 3-fold SUSY models realized with the choice
$f(z)=\rme^{\nu z}$. In the latter choice, the functions $A(z)$ in
(\ref{eq:Az}) and $Q(z)$ in (\ref{eq:BC}) read as
\begin{align}
A(z)=&\;\frac{c_{2}}{\nu}z\,\rme^{\nu z}-\frac{c_{2}+b_{2}\nu}{\nu^{2}}\,
 \rme^{\nu z}+\frac{c_{1}}{\nu}z^{2}-\frac{c_{1}+(b_{1}-c_{0})\nu}{\nu^{2}}
 z\notag\\
&-\frac{c_{0}-a_{2}+b_{0}\nu}{\nu^{2}}+\frac{a_{1}}{\nu^{2}}z\,\rme^{-\nu z}
 +\frac{a_{0}}{\nu^{2}}\,\rme^{-\nu z},
\label{eq:Az1}\\
Q(z)=&\;-\frac{c_{2}}{2}z\,\rme^{\nu z}+\frac{b_{2}}{2}\,\rme^{\nu z}-c_{1}
 z^{2}+\frac{c_{1}+(b_{1}-c_{0})\nu}{\nu}z\notag\\
&\;-\frac{c_{1}+(b_{1}-c_{0})\nu -2b_{0}\nu^{2}}{2\nu^{2}}-\frac{a_{1}}{2\nu}
 z\,\rme^{-\nu z}+\frac{a_{1}-a_{0}\nu}{2\nu^{2}}\,\rme^{-\nu z}.
\label{eq:Qz1}
\end{align}
The change of variable from $z$ to $q$ is determined by the first formula
in (\ref{eq:AQ}), but it cannot be integrated analytically in general.
Thus, we shall restrict ourselves to considering some particular cases
where the change of variable can be performed analytically. In what follows,
we shall present the change of variable $z(q)$, the functions $E(q)$ and
$F(q)$ in (\ref{eq:EF}) and $W(q)$ in (\ref{eq:AQ}), which altogether
characterize a type B model, the gauge factors $\cW_{3}^{\pm}(q)$ in
(\ref{eq:cW3+-}), the pair of potentials $V^{\pm}(q)$ in (\ref{eq:cond1}),
and the solvable sectors $\cV_{3}^{\pm}[q]$ in (\ref{eq:cV3}).\\

\noindent
\textbf{Example 1.}\quad $A(z)=2\alpha z.$\vspace{5pt}

This case is realized by putting $c_{2}=c_{1}=b_{2}=a_{1}=a_{0}=0$,
$a_{2}=c_{0}+b_{0}\nu$, and $c_{0}=b_{1}+2\alpha\nu$ in (\ref{eq:Az1}).
Then, we obtain the following:\vspace{5pt}

\noindent
\textit{Change of variable:}\quad $z(q)=\alpha q^{2}.$\vspace{5pt}\\
\textit{Functions:}
\begin{align}
E(q)=\frac{1}{q},\qquad F(q)=2\alpha\nu q,\qquad W(q)=\alpha\nu q
 -\frac{\alpha+b_{0}}{2\alpha q}.
\end{align}
\textit{Gauge factors:}
\begin{align}
\cW_{3}^{\pm}(q)=\frac{2\alpha\pm(\alpha+b_{0})}{2\alpha}\ln|q|
 \mp\frac{\alpha\nu}{2}q^{2}.
\end{align}
\textit{Potentials:}
\begin{align}
V^{\pm}(q)=\frac{\alpha^{2}\nu^{2}}{2}q^{2}+\frac{(b_{0}+\alpha
 \pm 2\alpha)(b_{0}+\alpha\pm 4\alpha)}{8\alpha^{2}q^{2}}
 -\frac{b_{0}+\alpha\mp 3\alpha}{6}\nu.
\end{align}
\textit{Solvable sectors:}
\begin{align}
\begin{split}
\cV_{3}^{-}[q]&=\braket{1, q^{2}, \rme^{\alpha\nu q^{2}}}q^{
 \frac{b_{0}-\alpha}{2\alpha}}\rme^{-\frac{\alpha\nu}{2}q^{2}},\\
\cV_{3}^{+}[q]&=\braket{1, q^{2}, \rme^{-\alpha\nu q^{2}}}q^{
 -\frac{3\alpha+b_{0}}{2\alpha}}\rme^{\frac{\alpha\nu}{2}q^{2}}.
\end{split}
\label{eq:ss1}
\end{align}
Both of the potentials are radial harmonic oscillators, not only
quasi-solvable but also solvable, and in particular quite similar
to one of the known type C $\cN$-fold SUSY models (with $\cN=3$),
Case~1 in Ref.~\cite{GT05}. However, a precise comparison tells us
that they are slightly different from each other in parameter values.\\

\noindent
\textbf{Example 2.}\quad $A(z)=\alpha z^{2}/2.$\vspace{5pt}

This case is realized by putting $c_{2}=b_{2}=a_{1}=a_{0}=0$,
$a_{2}=c_{0}+b_{0}\nu$, and $c_{1}=(c_{0}-b_{1})\nu=\alpha\nu/2$ in
(\ref{eq:Az1}). Then, we obtain the following:\vspace{5pt}

\noindent
\textit{Change of variable:}\quad $z(q)=\rme^{\sqal q}.$\vspace{5pt}\\
\textit{Functions:}
\begin{align}
E(q)=\sqal,\quad F(q)=\sqal\nu\,\rme^{\sqal q},\quad W(q)=\frac{\sqal\nu}{2}
 \,\rme^{\sqal q}-\frac{b_{0}}{\sqal}\,\rme^{-\sqal q}.
\end{align}
\textit{Gauge factors:}
\begin{align}
\cW_{3}^{\pm}(q)=\sqal q\mp\frac{\nu}{2}\,\rme^{\sqal q}\mp
 \frac{b_{0}}{\alpha}\,\rme^{-\sqal q}.
\end{align}
\textit{Potentials:}
\begin{align}
V^{\pm}(q)=\frac{\alpha\nu^{2}}{8}\,\rme^{2\sqal q}+\frac{(b_{0})^{2}}{
 2\alpha}\,\rme^{-2\sqal q}\pm\frac{\alpha\nu}{4}\,\rme^{\sqal q}\pm
 \frac{3b_{0}}{2}\,\rme^{-\sqal q}+\frac{2\alpha-b_{0}\nu}{6}.
\end{align}
\textit{Solvable sectors:}
\begin{align}
\begin{split}
\cV_{3}^{-}[q]&=\braket{1, \rme^{\sqal q}, \rme^{\nu\,\rme^{\sqal q}}}
 \exp\left(-\sqal q-\frac{\nu}{2}\,\rme^{\sqal q}-\frac{b_{0}}{\alpha}
 \,\rme^{-\sqal q}\right),\\
\cV_{3}^{+}[q]&=\braket{1, \rme^{\sqal q}, \rme^{-\nu\,\rme^{\sqal q}}}
 \exp\left(-\sqal q+\frac{\nu}{2}\,\rme^{\sqal q}+\frac{b_{0}}{\alpha}
 \,\rme^{-\sqal q}\right).
\end{split}
\end{align}
This model quite resembles one of the type A $\cN$-fold SUSY systems
(with $\cN=3$), Case III in Ref.~\cite{Ta09}, which is a quasi-solvable
generalization of Morse potential, but actually has a slight difference
between them in the precise potential form. The solvable sectors are
also different.\\

\noindent
\textbf{Example 3.}\quad $A(z)=(\alpha\,\rme^{\frac{\nu}{2}z}+\beta\,
\rme^{-\frac{\nu}{2}z})^{2}/2.$\vspace{5pt}

This case is realized by putting $c_{2}=c_{1}=a_{1}=0$, $c_{0}=b_{1}$,
$(b_{0}\nu+c_{0}-a_{2})^{2}=-4a_{0}b_{2}\nu$, $2b_{2}=-\alpha^{2}\nu$,
and $2a_{0}=\beta^{2}\nu^{2}$ in (\ref{eq:Az1}). For convenience, we
assume that $\alpha\beta>0$. Then, we obtain the following:\vspace{5pt}

\noindent
\textit{Change of variable:}\quad $z(q)=\frac{2}{\nu}\ln\tan\frac{
\sqab\nu}{2}q.$\vspace{5pt}\\
\textit{Functions:}
\begin{align}
\begin{split}
&E(q)=-\sqab\nu\frac{\cos\sqab\nu q}{\sin\sqab\nu q},\quad F(q)=\frac{
 2\sqab\nu}{\sin\sqab\nu q},\\
&W(q)=\frac{\sqab\nu}{2\sin\sqab\nu q}
 -\frac{2b_{0}+\alpha\beta\nu}{4\sqab}\sin\sqab\nu q.
\end{split}
\end{align}
\textit{Gauge factors:}
\begin{align}
\cW_{3}^{\pm}(q)=-\ln\left|\sin\sqab\nu q\right|\mp\frac{1}{2}\ln\left|
 \tan\frac{\sqab\nu}{2}q\right|\mp\frac{2b_{0}+\alpha\beta\nu}{
 4\alpha\beta\nu}\cos\sqab\nu q.
\end{align}
\textit{Potentials:}
\begin{align}
V^{\pm}(q)=&\;\frac{(2b_{0}+\alpha\beta\nu)^{2}}{32\alpha\beta}\sin^{2}
 \sqab\nu q+\frac{\alpha\beta\nu^{2}}{8\sin^{2}\sqab\nu q}+\frac{2b_{0}
 -7\alpha\beta\nu}{24}\nu\notag\\
&\;\mp\frac{3(2b_{0}+\alpha\beta\nu)\nu}{8}\cos\sqab\nu q\pm\frac{
 \alpha\beta\nu^{2}}{4}\frac{\cos\sqab\nu q}{\sin^{2}\sqab\nu q}.
\end{align}
\textit{Solvable sectors:}
\begin{align}
\begin{split}
\cV_{3}^{-}[q]=&\;\braket{1, \ln\tan\frac{\sqab\nu}{2}q, \tan^{2}\frac{
 \sqab\nu}{2}q}\left(\sin\frac{\sqab\nu}{2}q\right)^{1/2}\\
&\;\times\left(\cos\frac{\sqab\nu}{2}q\right)^{3/2}\exp\left(
 -\frac{2b_{0}+\alpha\beta\nu}{4\alpha\beta\nu}\cos\sqab\nu q\right),\\
\cV_{3}^{+}[q]=&\;\braket{1, \ln\tan\frac{\sqab\nu}{2}q, \tan^{-2}\frac{
 \sqab\nu}{2}q}\left(\sin\frac{\sqab\nu}{2}q\right)^{3/2}\\
&\;\times\left(\cos\frac{\sqab\nu}{2}q\right)^{1/2}\exp\left(
 \frac{2b_{0}+\alpha\beta\nu}{4\alpha\beta\nu}\cos\sqab\nu q\right).
\end{split}
\end{align}
As in the previous example, this model also resembles one of the type A
$\cN$-fold SUSY systems (with $\cN=3$), Case IV in Ref.~\cite{Ta09},
which is a quasi-solvable generalization of P\"{o}schl--Teller potential,
if the trigonometric functions in the above are properly replaced by
hyperbolic ones. See also Ref.~\cite{GKO93}.
But the element $\ln\tan\sqrt{\alpha\beta}\nu q/2$
in the solvable sectors clearly shows the peculiarity of type B, which
does not appear in any type A system.

\section{Type A, B, and C 3-fold SUSY with Monomial Subspaces}
\label{sec:ABC}

In this section, we shall examine some particular cases of the most general
type B $3$-fold SUSY investigated in Section~\ref{sec:B3}. Virtually all the
$\cN$-fold SUSY models (including the ones called with other terminologies)
so far constructed in the literature are of the types which preserve a monomial
or polynomial subspace. The latter types of second-order linear differential
operators for the monomial cases were first classified in Ref.~\cite{PT95}.
According to it and to the later reconsiderations in Refs.~\cite{GT04,GT05},
there are essentially three inequivalent monomial subspaces, except for a few
low-dimensional ones, preserved by second-order linear differential operators.
They are respectively called type A, type B, and type C monomial
subspaces~\cite{GT05,Ta09}. $\cN$-fold SUSY systems associated with them were
investigated in general fashions in Refs.~\cite{AST01a,ANST01,Ta03a} for type
A, in Ref.~\cite{GT04} for type B, and in Ref.~\cite{GT05} for type C monomial
subspaces.

As we have shown in Section~\ref{sec:B3}, any second-order linear differential
operator which preserves a three-dimensional linear space of functions
admitting an analytic expression has type B $3$-fold SUSY. It means that all 
the aforementioned systems which preserve a monomial or polynomial subspace 
must have type B $3$-fold SUSY too when $\cN=3$. Hence, it is interesting to 
know how those systems in the literature would be realized in the most general 
type B $3$-fold SUSY obtained in Section~\ref{sec:B3}.

Let us first begin with type C, since, as we will see later, type A and type B 
can be regarded as special cases of type C. A type C monomial subspace 
$\tcV_{\cN_{1},\,\cN_{2}}^{(\rmC)}$ of dimension $\cN$ is defined
by~\cite{GT05}
\begin{align}
\tcV_{\cN_{1},\,\cN_{2}}^{(\rmC)}=\braket{1,z,\dots,z^{\cN_{1}-1},z^{\lambda},
 z^{\lambda+1},\dots,z^{\lambda+\cN_{2}-1}},\qquad \cN=\cN_{1}+\cN_{2}\ge 3,
\label{eq:typeC}
\end{align}
where $\cN_{1}$ and $\cN_{2}$ are positive integers and $\lambda$ is a real
number.
The linear space (\ref{eq:tV3-}) we have considered reduces to a type C
monomial subspace of dimension $3$ with $\cN_{1}=2$ and $\cN_{2}=1$ when 
$f(z)=z^{\lambda}$:
\begin{align}
\tcV_{3}^{-}\bigr|_{f(z)=z^{\lambda}}=\braket{1,z,z^{\lambda}}=
 \tcV_{2,1}^{(\rmC)}\qquad (\lambda\neq -2,-1,\dots,3).
\label{eq:tV3-C}
\end{align}
The corresponding gauged $3$-fold supercharge component (\ref{eq:tP3-})
which annihilates this space (\ref{eq:tV3-C}) reads as
\begin{align}
\tP_{3}^{-}=z'(q)^{3}\left( \frac{\rmd^{2}}{\rmd z^{2}}-\frac{\lambda-2}{z}
 \right)\frac{\rmd^{2}}{\rmd z^{2}},\qquad \tw_{2}^{[3]}(z)=
 -\frac{\lambda-2}{z},
\end{align}
where $\tw_{2}^{[3]}(z)$ is calculated with (\ref{eq:tw2}). The set of eight 
second-order linear differential operators (\ref{eq:Js}) which preserve the 
three-dimensional type C monomial subspace (\ref{eq:tV3-C}) is
\begin{align}
\begin{split}
&\lambda(\lambda-1) J_{1}:=\tJ_{1}(\lambda)=z^{2-\lambda}\frac{\rmd^{2}}{
 \rmd z^{2}},\qquad \lambda(\lambda-1) J_{2}:=\tJ_{2}(\lambda)=z^{3-\lambda}
 \frac{\rmd^{2}}{\rmd z^{2}},\\
&\lambda(\lambda-1) J_{3}:=\tJ_{3}(\lambda)=z^{2}\frac{\rmd^{2}}{\rmd z^{2}},
 \qquad (\lambda-1) J_{4}:=\tJ_{4}(\lambda)=z\frac{\rmd^{2}}{\rmd z^{2}}
 -(\lambda-1)\frac{\rmd}{\rmd z},\\
&(\lambda-1) J_{5}:=\tJ_{5}(\lambda)=z^{2}\frac{\rmd^{2}}{\rmd z^{2}}
 -(\lambda-1)z\frac{\rmd}{\rmd z},\\
&(\lambda-1) J_{6}:=\tJ_{6}(\lambda)=z^{\lambda+1}\frac{\rmd^{2}}{\rmd z^{2}}
 -(\lambda-1)z^{\lambda}\frac{\rmd}{\rmd z},\\
&\lambda J_{7}:=\tJ_{7}(\lambda)=z^{3}\frac{\rmd^{2}}{\rmd z^{2}}-\lambda
 z^{2}\frac{\rmd}{\rmd z}+\lambda z,\\
&\lambda J_{8}:=\tJ_{8}(\lambda)=z^{\lambda+2}\frac{\rmd^{2}}{\rmd z^{2}}
 -\lambda z^{\lambda+1}\frac{\rmd}{\rmd z}+\lambda z^{\lambda}.
\end{split}
\label{eq:JsC}
\end{align}
On the other hand, in the existing literature there are six linearly
independent linear differential operators of at most second order preserving
the linear space (\ref{eq:tV3-C}), which are summarized in (\ref{eq:ApJsC})
and (\ref{eq:ApJsC'}). The correspondence between (\ref{eq:JsC}) and the
latter is as follows:
\begin{align}
\begin{split}
(\lambda-1) J_{0}^{(\rmC)}(\lambda)=\tJ_{3}(\lambda)-\tJ_{5}(\lambda),\quad
 J_{0-}^{(\rmC)}(\lambda)=\tJ_{4}(\lambda),\quad
 J_{00}^{(\rmC)}(\lambda)=\tJ_{3}(\lambda),\\
 J_{+0}^{(\rmC)}(\lambda)=\tJ_{7}(\lambda),\qquad
 J_{\# -}^{(\rmC)}(\lambda)=\tJ_{6}(\lambda),\qquad
 J_{\# 0}^{(\rmC)}(\lambda)=\tJ_{8}(\lambda).
\end{split}
\label{eq:reJsC}
\end{align}
Hence, we see that the two operators $\tJ_{1}(\lambda)$ and
$\tJ_{2}(\lambda)$ have been missed so far in the literature. By using
(\ref{eq:JsC}) and (\ref{eq:reJsC}), the most general second-order
linear differential operator (\ref{eq:tH-2}) having an invariant 
type C monomial subspace (\ref{eq:tV3-C}) is expressed as
\begin{align}
\tH^{-}=&\;-\frac{c_{2}}{\lambda}\tJ_{8}(\lambda)+\frac{b_{2}}{\lambda-1}
 \tJ_{6}(\lambda)-\frac{a_{1}}{\lambda(\lambda-1)}\tJ_{2}(\lambda)
 -\frac{a_{0}}{\lambda(\lambda-1)}\tJ_{1}(\lambda)\notag\\
&\;-\ta_{3}J_{+0}^{(\rmC)}(\lambda)-\ta_{2}J_{00}^{(\rmC)}(\lambda)
 -\ta_{1}J_{0-}^{(\rmC)}(\lambda)-\tb_{0}J_{0}^{(\rmC)}(\lambda)-c_{0},
\end{align}
where
\begin{align}
\ta_{3}=\frac{c_{1}}{\lambda},\quad \ta_{2}=-\frac{b_{1}-c_{0}}{\lambda-1}
 +\frac{a_{2}-c_{0}}{\lambda(\lambda-1)},\quad \ta_{1}=-\frac{b_{0}}{
 \lambda-1},\quad \tb_{0}=b_{1}-c_{0}.
\label{eq:paraC}
\end{align}
When we put $c_{2}=b_{2}=a_{1}=a_{0}=0$, it reduces to the gauged type C
$3$-fold SUSY Hamiltonian constructed in Eq.~(3.8) of Ref.~\cite{GT05}.

The gauged $3$-fold superchage component (\ref{eq:bP3+}) and the solvable
sector (\ref{eq:bV3+}) annihilated by it for the plus component in this
case read as
\begin{align}
\bar{P}_{3}^{+}=-z'(q)^{3}\frac{\rmd^{2}}{\rmd z^{2}}\left(
 \frac{\rmd}{\rmd z}+\frac{\lambda-2}{z}\right),\qquad \bar{\cV}_{3}^{+}
 =z\braket{1,z,z^{1-\lambda}}.
\label{eq:bV3+C}
\end{align}
The set of eight linearly independent second-order linear differential
operators (\ref{eq:Ks}) which preserve the latter space reduces to
\begin{align}
\begin{split}
&\lambda(\lambda-1) K_{1}:=\tK_{1}(\lambda)=z^{2-\lambda}\frac{\rmd^{2}}{
 \rmd z^{2}}+(\lambda-2) z^{1-\lambda}\frac{\rmd}{\rmd z}-(\lambda-2)
 z^{-\lambda},\\
&\lambda(\lambda-1) K_{2}:=\tK_{2}(\lambda)=z^{3-\lambda}\frac{\rmd^{2}}{
 \rmd z^{2}}+(\lambda-3) z^{2-\lambda}\frac{\rmd}{\rmd z}-2(\lambda-2)
 z^{1-\lambda},\\
&\lambda(\lambda-1) K_{3}:=\tK_{3}(\lambda)=z^{2}\frac{\rmd^{2}}{\rmd z^{2}}
 -2 z\frac{\rmd}{\rmd z}+2,\\
&(\lambda-1) K_{4}:=\tK_{4}(\lambda)=z\frac{\rmd^{2}}{\rmd z^{2}}+(\lambda-2)
 \frac{\rmd}{\rmd z}-(\lambda-2) z^{-1},\\
&(\lambda-1) K_{5}:=\tK_{5}(\lambda)=z^{2}\frac{\rmd^{2}}{\rmd z^{2}}
 +(\lambda-3) z\frac{\rmd}{\rmd z}-2(\lambda-2),\\
&(\lambda-1) K_{6}:=\tK_{6}(\lambda)=z^{\lambda+1}\frac{\rmd^{2}}{\rmd z^{2}}
 -2 z^{\lambda}\frac{\rmd}{\rmd z}+2 z^{\lambda-1},\\
&\lambda K_{7}:=\tK_{7}(\lambda)=z^{3}\frac{\rmd^{2}}{\rmd z^{2}}+(\lambda-3)
 z^{2}\frac{\rmd}{\rmd z}-2(\lambda-2) z,\\
&\lambda K_{8}:=\tK_{8}(\lambda)=z^{\lambda+2}\frac{\rmd^{2}}{\rmd z^{2}}
 -2 z^{\lambda+1}\frac{\rmd}{\rmd z}+2 z^{\lambda}.
\end{split}
\end{align}
The relation to
the known quasi-solvable operators listed in (\ref{eq:ApKsC}) is as follows:
\begin{align}
\begin{split}
(\lambda-1) K_{0}^{(\rmC)}(\lambda)=\tK_{5}(\lambda)-\tK_{3}(\lambda)
 +\lambda-1,\qquad K_{0-}^{(\rmC)}(\lambda)=\tK_{4}(\lambda),\\
K_{00}^{(\rmC)}(\lambda)=\tK_{3}(\lambda),\ \ 
 K_{+0}^{(\rmC)}(\lambda)=\tK_{7}(\lambda),\ \ 
 K_{\# -}^{(\rmC)}(\lambda)=\tK_{1}(\lambda),\ \ 
 K_{\# 0}^{(\rmC)}(\lambda)=\tK_{2}(\lambda).
\end{split}
\label{eq:reKsC}
\end{align}
Hence, we see again that the two operators $\tK_{6}(\lambda)$ and 
$\tK_{8}(\lambda)$ have been missed in the literature. 

Finally, we note that the function $F(q)$ defined in (\ref{eq:EF}) is
calculated in this case as
\begin{align}
F(q)=\frac{(\lambda-2) z'(q)}{z(q)},
\end{align}
and coincides (up to a multiplicative factor) with the \textit{ad hoc}
constraint $F(q)=z'(q)/z(q)$ made in Ref.~\cite{GT05}, Eq.~(3.17).
We now understand that it naturally comes from the relation (\ref{eq:EF})
for the general case of an arbitrary $f(z)$.\\

Let us next consider the case associated with a type B monomial subspace. An
$\cN$-dimensional type B monomial space is defined by~\cite{GT04,GT05}
\begin{align}
\tcV_{\cN}^{(\rmB)}=\braket{1,z,\dots,z^{\cN-2},z^{\cN}}.
\label{eq:typeB}
\end{align}
For $\cN=3$, it can be realized as a special case of the linear space
(\ref{eq:tV3-}) with $f(z)=z^{3}$ or as a special case of the type C monomial 
space (\ref{eq:tV3-C}) with $\lambda=3$:
\begin{align}
\tcV_{3}^{-}\bigr|_{f(z)=z^{3}}=\tcV_{2,1}^{(\rmC)}\bigr|_{\lambda=3}
 =\braket{1,z,z^{3}}=\tcV_{3}^{(\rmB)}.
\label{eq:tV3-B}
\end{align}
The corresponding gauged $3$-fold supercharge component (\ref{eq:tP3-}) which
annihilates this space (\ref{eq:tV3-B}) reads as
\begin{align}
\tP_{3}^{-}=z'(q)^{3}\left( \frac{\rmd}{\rmd z}-\frac{1}{z}\right)
 \frac{\rmd^{2}}{\rmd z^{2}},\qquad \tw_{2}^{[3]}(z)=-\frac{1}{z},
\end{align}
where $\tw_{2}^{[3]}(z)$ is calculated with (\ref{eq:tw2}). The set of eight
second-order linear differential operators (\ref{eq:Js}) which leave the 
three-dimensional type B monomial subspace (\ref{eq:tV3-B}) invariant is
\begin{align}
\begin{split}
&\tJ_{1}(3)=z^{-1}\frac{\rmd^{2}}{\rmd z^{2}},\qquad \tJ_{2}(3)=\frac{
 \rmd^{2}}{\rmd z^{2}},\qquad \tJ_{3}(3)=z^{2}\frac{\rmd^{2}}{\rmd z^{2}},\\
&\tJ_{4}(3)=z\frac{\rmd^{2}}{\rmd z^{2}}-2\frac{\rmd}{\rmd z},\quad
 \tJ_{5}(3)=z^{2}\frac{\rmd^{2}}{\rmd z^{2}}-2 z\frac{\rmd}{\rmd z},\quad
 \tJ_{6}(3)=z^{4}\frac{\rmd^{2}}{\rmd z^{2}}-2 z^{3}\frac{\rmd}{\rmd z},\\
&\tJ_{7}(3)=z^{3}\frac{\rmd^{2}}{\rmd z^{2}}-3 z^{2}\frac{\rmd}{\rmd z}+3
 z,\qquad \tJ_{8}(3)=z^{5}\frac{\rmd^{2}}{\rmd z^{2}}-3 z^{4}
 \frac{\rmd}{\rmd z}+3 z^{3}.
\end{split}
\label{eq:JsB}
\end{align}
On the other hand, in the existing literature there are seven linearly 
independent linear differential operators of at most second order preserving 
the linear space (\ref{eq:tV3-B}), which are summarized in (\ref{eq:ApJsB})
and (\ref{eq:ApJsB'}).
The correspondence between (\ref{eq:JsB}) and the latter is as follows:
\begin{align}
\begin{split}
&2 J_{0}^{(\rmB)}=\tJ_{3}(3)-\tJ_{5}(3),\qquad J_{--}^{(\rmB)}=\tJ_{2}(3),
 \qquad J_{0-}^{(\rmB)}=\tJ_{4}(3),\\
&J_{00}^{(\rmB)}=\tJ_{3}(3),\quad J_{+0}^{(\rmB)}=\tJ_{7}(3),\quad
 J_{++}^{(\rmB)}=\tJ_{6}(3),\quad J_{3+}^{(\rmB)}=\tJ_{8}(3).
\end{split}
\label{eq:reJsB}
\end{align}
Hence, we see that the one operator $\tJ_{1}(3)$ has been missed so far
in the literature. By using (\ref{eq:JsB}) and (\ref{eq:reJsB}), the most
general second-order linear differential operator (\ref{eq:tH-2}) having
an invariant type B monomial subspace (\ref{eq:tV3-B}) is expressed as
\begin{align}
\tH^{-}=-\frac{c_{2}}{3}\tJ_{8}(3)-\frac{a_{0}}{6}\tJ_{1}(3)-\sum_{
 \substack{i,j=+,0,-\\ i\geq j}}
 \ta_{ij}\tJ_{ij}^{(\rmB)}-\tb_{0}\tJ_{0}^{(\rmB)}-c_{0},
\end{align}
where
\begin{align}
\begin{split}
&2\ta_{++}=-b_{2},\qquad 3\ta_{+0}=c_{1},\qquad 6\ta_{00}=a_{2}-3 b_{1}
 -2 c_{0},\\
&2\ta_{0-}=-b_{0},\qquad 6\ta_{--}=a_{1},\qquad \tb_{0}=b_{1}-c_{0}.
\end{split}
\label{eq:paraB}
\end{align}
When we put $c_{2}=a_{0}=0$, it reduces to the gauged type B $3$-fold
SUSY Hamiltonian in Eq.~(3.12) of Ref.~\cite{GT04}. 

The gauged $3$-fold superchage component (\ref{eq:bP3+}) and the solvable
sector (\ref{eq:bV3+}) annihilated by it for the plus component in this case 
read as
\begin{align}
\bar{P}_{3}^{+}=-z'(q)^{3}\frac{\rmd^{2}}{\rmd z^{2}}\left(\frac{\rmd}{\rmd z}
 +\frac{1}{z}\right),\qquad \bar{\cV}_{3}^{+}=z^{-1}\braket{1,z^{2},z^{3}}.
\label{eq:bV3+B}
\end{align}
The set of eight linearly independent second-order linear differential
operators (\ref{eq:Ks}) which preserve the latter space reduces to
\begin{align}
\begin{split}
&\tK_{1}(3)=z^{-1}\frac{\rmd^{2}}{\rmd z^{2}}+z^{-2}\frac{\rmd}{\rmd z}
 -z^{-3},\qquad \tK_{2}(3)=\frac{\rmd^{2}}{\rmd z^{2}}-2 z^{-2},\\
&\tK_{3}(3)= z^{2}\frac{\rmd^{2}}{\rmd z^{2}}-2 z\frac{\rmd}{\rmd z}+2,\qquad
 \tK_{4}(3)=z\frac{\rmd^{2}}{\rmd z^{2}}+\frac{\rmd}{\rmd z}-z^{-1},\\
&\tK_{5}(3)=z^{2}\frac{\rmd^{2}}{\rmd z^{2}}+2,\qquad \tK_{6}(3)=z^{4}
 \frac{\rmd^{2}}{\rmd z^{2}}-2 z^{3}\frac{\rmd}{\rmd z}+2 z^{2},\\
&\tK_{7}(3)=z^{3}\frac{\rmd^{2}}{\rmd z^{2}}-2 z,\qquad \tK_{8}(3)=z^{5}
 \frac{\rmd^{2}}{\rmd z^{2}}-2 z^{4}\frac{\rmd}{\rmd z}+2 z^{3}.
\end{split}
\end{align}
The relation to the known quasi-solvable operators listed
in (\ref{eq:ApKsB}) and (\ref{eq:ApKsB'}) is as follows:
\begin{align}
\begin{split}
&2 K_{0}^{(\rmB)}=\tK_{5}(3)-\tK_{3}(3),\qquad K_{--}^{(\rmB)}=\tK_{2}(3),
 \qquad K_{0-}^{(\rmB)}=\tK_{4}(3),\\
&K_{00}^{(\rmB)}=\tK_{5}(3)-2,\quad K_{+0}^{(\rmB)}=\tK_{7}(3),\quad
 K_{++}^{(\rmB)}=\tK_{6}(3),\quad K_{3+}^{(\rmB)}=\tK_{8}(3).
\end{split}
\end{align}
Hence, we see again that the one operator $\tK_{1}(3)$ has been
missed in the literature.

Finally, the function $F(q)$ defined in (\ref{eq:EF}) is calculated in
this case as $F(q)=z'(q)/z(q)$, and coincides exactly with the \textit{ad hoc}
constraint on $F(q)$ made in Eq.~(3.6) of Ref.~\cite{GT04}. As in the case
of type C, it naturally comes from the general relation (\ref{eq:EF}).\\

In the last, we shall consider the case associated with a type A monomial 
subspace. An $\cN$-dimensional type A monomial space is defined
by~\cite{ANST01,GT05}
\begin{align}
\tcV_{\cN}^{(\rmA)}=\braket{1,z,\dots,z^{\cN-1}}.
\label{eq:typeA}
\end{align}
For $\cN=3$, it can be realized as a special case of the linear space
(\ref{eq:tV3-}) with $f(z)=z^{2}$ or as a special case of the type C
monomial space (\ref{eq:tV3-C}) with $\lambda=2$:
\begin{align}
\tcV_{3}^{-}\bigr|_{f(z)=z^{2}}=\tcV_{2,1}^{(\rmC)}\bigr|_{\lambda=2}
 =\braket{1,z,z^{2}}=\tcV_{3}^{(\rmA)}.
\label{eq:tV3-A}
\end{align}
The corresponding gauged $3$-fold supercharge component (\ref{eq:tP3-})
which annihilates this space (\ref{eq:tV3-A}) reads as
\begin{align}
\tP_{3}^{-}=z'(q)^{3}\frac{\rmd^{3}}{\rmd z^{3}},\qquad \tw_{2}^{[3]}(z)=0,
\label{eq:tP3-A}
\end{align}
where $\tw_{2}^{[3]}(z)$ is calculated with (\ref{eq:tw2}). The set of eight 
second-order linear differential operators (\ref{eq:Js}) which preserve the 
three-dimensional type A monomial subspace (\ref{eq:tV3-A}) is
\begin{align}
\begin{split}
&\tJ_{1}(2)=\frac{\rmd^{2}}{\rmd z^{2}},\qquad \tJ_{2}(2)=z\frac{\rmd^{2}}{
 \rmd z^{2}}\qquad \tJ_{3}(2)=z^{2}\frac{\rmd^{2}}{\rmd z^{2}},\\
&\tJ_{4}(2)=z\frac{\rmd^{2}}{\rmd z^{2}}-\frac{\rmd}{\rmd z},\quad
 \tJ_{5}(2)=z^{2}\frac{\rmd^{2}}{\rmd z^{2}}-z\frac{\rmd}{\rmd z},\quad
 \tJ_{6}(2)=z^{3}\frac{\rmd^{2}}{\rmd z^{2}}-z^{2}\frac{\rmd}{\rmd z},\\
&\tJ_{7}(2)=z^{3}\frac{\rmd^{2}}{\rmd z^{2}}-2 z^{2}\frac{\rmd}{\rmd z}+2
 z,\qquad \tJ_{8}(2)=z^{4}\frac{\rmd^{2}}{\rmd z^{2}}-2 z^{3}\frac{\rmd}{
 \rmd z}+2 z^{2}.
\end{split}
\label{eq:JsA}
\end{align}
In contrast with the previous two cases of type C and type B, it is entirely
equivalent with the set of eight linearly independent linear differential 
operators of at most second order preserving the linear space
(\ref{eq:tV3-A}) already appeared in the existing literature, which are 
summarized in (\ref{eq:ApJsA}). The correspondence between (\ref{eq:JsA})
and (\ref{eq:ApJsA}) is as follows:
\begin{align}
\begin{split}
&J_{-}^{(\rmA)}=\tJ_{2}(2)-\tJ_{4}(2),\qquad J_{0}^{(\rmA)}=\tJ_{3}(2)
 -\tJ_{5}(2),\qquad J_{+}^{(\rmA)}=\tJ_{6}(2)-\tJ_{7}(2),\\
&J_{--}^{(\rmA)}=\tJ_{1}(2),\quad J_{0-}^{(\rmA)}=\tJ_{2}(2),\quad
 J_{00}^{(\rmA)}=\tJ_{3}(3),\quad J_{+0}^{(\rmA)}=\tJ_{6}(3),\quad
 J_{++}^{(\rmA)}=\tJ_{8}(3).
\end{split}
\label{eq:reJsA}
\end{align}
By using (\ref{eq:JsA}) and (\ref{eq:reJsA}), the most general second-order 
linear differential operator (\ref{eq:tH-2}) having an invariant type A 
monomial subspace (\ref{eq:tV3-A}) is expressed as
\begin{align}
\tH^{-}=-\sum_{\substack{i,j=+,0,-\\ i\geq j}}\ta_{ij}J_{ij}^{(\rmA)}
 +\sum_{i=+,0,-}\tb_{i}J_{i}^{(\rmA)}-c_{0},
\end{align}
where
\begin{align}
\begin{split}
&2\ta_{++}=c_{2},\qquad 2\ta_{+0}=-2 b_{2}+c_{1},\qquad 2\ta_{00}=a_{2}
 -2 b_{1}+c_{0},\\
&2\ta_{0-}=a_{1}-2 b_{0},\quad 2\ta_{--}=a_{0},\quad 2\tb_{+}=c_{1},\quad
 \tb_{0}=-b_{1}+c_{0},\quad \tb_{-}=-b_{0}.
\end{split}
\label{eq:paraA}
\end{align}
It exactly coincides with the gauged type A $3$-fold Hamiltonian in
Eq.~(4.22) of Ref.~\cite{ANST01}, which is originated from the $\fsl(2)$ 
Lie-algebraic quasi-solvable operators in Ref.~\cite{Tu88}.

The gauged $3$-fold superchage component (\ref{eq:bP3+}) and the solvable
sector (\ref{eq:bV3+}) annihilated by it for the plus component in this case 
read as
\begin{align}
\bar{P}_{3}^{+}=-z'(q)^{3}\frac{\rmd^{3}}{\rmd z^{3}},\qquad \bar{\cV}_{3}^{+}=
 \braket{1,z,z^{2}},
\end{align}
and are entirely identical with the ones for the minus component,
(\ref{eq:tV3-A}) and (\ref{eq:tP3-A}). The set of eight linearly independent
second-order linear differential operators (\ref{eq:Ks}) which preserve
the latter space reduces to
\begin{align}
\begin{split}
&\tK_{1}(2)=\frac{\rmd^{2}}{\rmd z^{2}},\qquad \tK_{2}(2)=z\frac{\rmd^{2}}{
 \rmd z^{2}}-\frac{\rmd}{\rmd z},\qquad \tK_{3}(2)=z^{2}\frac{\rmd^{2}}{
 \rmd z^{2}}-2 z\frac{\rmd}{\rmd z}+2,\\
&\tK_{4}(2)=z\frac{\rmd^{2}}{\rmd z^{2}},\quad \tK_{5}(2)=z^{2}
 \frac{\rmd^{2}}{\rmd z^{2}}-z\frac{\rmd}{\rmd z},\quad \tK_{6}(2)=z^{3}
 \frac{\rmd^{2}}{\rmd z^{2}}-2 z^{2}\frac{\rmd}{\rmd z}+2 z,\\
&\tK_{7}(2)=z^{3}\frac{\rmd^{2}}{\rmd z^{2}}-z^{2}\frac{\rmd}{\rmd z},\qquad
 \tK_{8}(2)=z^{4}\frac{\rmd^{2}}{\rmd z^{2}}-2 z^{3}\frac{\rmd}{\rmd z}+2
 z^{2},
\end{split}
\end{align}
and is again equivalent with the set of eight linearly independent linear 
differential operators of at most second order preserving the linear space
(\ref{eq:tV3-A}) in (\ref{eq:ApJsA}). The correspondence between
(\ref{eq:JsA}) and (\ref{eq:ApJsA}) is as follows (note that 
$K_{i}^{(\rmA)}=J_{i}^{(\rmA)}$ and $K_{ij}^{(\rmA)}=J_{ij}^{(\rmA)}$
($i,j=+,0,-$) in type A):
\begin{align}
\begin{split}
&K_{-}^{(\rmA)}=\tK_{4}(2)-\tK_{2}(2),\qquad K_{0}^{(\rmA)}=\tK_{5}(2)
 -\tK_{3}(2)+2,\\
&K_{+}^{(\rmA)}=\tK_{7}(2)-\tK_{6}(2),\qquad K_{--}^{(\rmA)}=\tK_{1}(2),
 \qquad K_{0-}^{(\rmA)}=\tK_{4}(2),\\
&K_{00}^{(\rmA)}=2\tK_{5}(2)-\tK_{3}(2)+2,\qquad K_{+0}^{(\rmA)}=\tK_{7}(2),
 \qquad K_{++}^{(\rmA)}=\tK_{8}(2).
\end{split}
\end{align}
Finally, the function $F(q)$ defined in (\ref{eq:EF}) vanishes in this case,
$F(q)=0$, which naturally explains the reason why type A $3$-fold SUSY can
be characterized only by two functions $E(q)$ and $W(q)$.

\section{Commutation Relations of the Operators}
\label{sec:CR}

It is well known that some quasi-solvable operators are constructed by
enveloping algebra of a Lie algebra. In fact, the discovery of the
underlying enveloping algebra of $\fsl(2)$ in \cite{Tu88} promoted
the various Lie-algebraic attempts to construct new quasi-solvable models,
see, e.g., \cite{GKO94b,Tu94} and references cited therein.
However, recent development has shown that there exist several
quasi-solvable systems which are not expressible in terms of a
differential operator representation of a particular Lie algebra
\cite{GT05,GT04,PT95,GKM05,GKM07}. The partial success
of the Lie-algebraic approach relies on the elementary fact that, if two
operators $J_{i}$ and $J_{j}$ ($i\neq j$) preserve a linear space $\tcV$,
so does its commutator:
\begin{align}
J_{i}\tcV\subset\tcV,\ J_{j}\tcV\subset\tcV\quad\Longrightarrow\quad
 [J_{i},J_{j}]\tcV\subset\tcV.
\end{align}
Needless to say, however, the commutator $[J_{i},J_{j}]$ usually has
higher order in derivatives than $J_{i}$ and $J_{j}$, and as a consequence
is not necessarily included in the set of quasi-solvable operators which 
is originally considered. In other words, Lie algebra is in general not
closed finitely in a finite set of quasi-solvable operators.

In our present case, each commutation relation of $J_{i}$ ($i=1,\dots,8$)
is indeed a linear differential operator of third order and thus cannot be
expressed as a linear combination of the second-order $J_{i}$, except for
the cases when something particular takes place. For a reference, we
list every commutation relation in Appendix~\ref{app:CR}.
To see what could happen when our quasi-solvable operators turn to be
Lie-algebraic, let us consider the following three linear combinations
of the operators $J_{i}$:
\begin{align}
J_{-}=J_{2}+\alpha_{-}J_{4},\quad J_{0}=J_{3}+\alpha_{0}J_{5},\quad
 J_{+}=J_{6}+\alpha_{+}J_{7},
\end{align}
where $\alpha_{i}$ ($i=-,0,+$) are constants. Their commutators are
calculated as
\begin{align}
[J_{-},J_{0}]=&\;2\frac{zf'-f+\alpha_{0}z(zf''-f')-\alpha_{-}[ff''-(f')^{2}]
 }{f''}\frac{\rmd}{\rmd z}J_{1}\notag\\
&+(\alpha_{0}+1)J_{2}+2\alpha_{0}\frac{z+\alpha_{-}f'}{f''}
 \frac{\rmd}{\rmd z}J_{4}-\alpha_{-}\alpha_{0}J_{4},
\label{eq:com1}\\
[J_{+},J_{0}]=&-2\frac{f^{2}f''-f(f')^{2}+\alpha_{+}z[(\alpha_{0}+1)zff''
 -z(f')^{2}+ff']}{f''}\frac{\rmd}{\rmd z}J_{1}\notag\\
&-2\frac{ff'+\alpha_{0}(zf'-f)f'-\alpha_{+}\alpha_{0}z(zf'-f)}{f''}
 \frac{\rmd}{\rmd z}J_{4}-(\alpha_{0}+1)J_{6}\notag\\
&-2\alpha_{+}\frac{f+\alpha_{0}zf'}{f''}\frac{\rmd}{\rmd z}J_{9}\
+\alpha_{+}\alpha_{0}J_{7},\\
[J_{+},J_{-}]=&-2\frac{f(zf''-f')+\alpha_{+}z(z^{2}f''-zf'+f)+\alpha_{+}
 \alpha_{-}zff''}{f''}\frac{\rmd}{\rmd z}J_{1}\notag\\
&-(\alpha_{+}\alpha_{-}+1)J_{3}-2\frac{(z+\alpha_{-}f')f'}{f''}
 \frac{\rmd}{\rmd z}J_{4}+(2\alpha_{+}\alpha_{-}-1)zJ_{4}\notag\\
&-2\alpha_{+}\frac{z+\alpha_{-}f'}{f''}
 \frac{\rmd}{\rmd z}J_{9}+\alpha_{+}\alpha_{-}.
\label{eq:com3}
\end{align}
Hence, it is necessary, though not sufficient, for them to satisfy
\begin{align}
z+\alpha_{-}f'(z)=f(z)+\alpha_{0}zf'(z)=0,
\end{align}
in order that they are of second order. This necessary condition is easily
solved as
\begin{align}
\alpha_{0}=-\frac{1}{2},\qquad f(z)=-\frac{z^{2}}{2\alpha_{-}}.
\label{eq:nc1}
\end{align}
The second formula means that it can happen only in the case of type A.
In the latter case, the commutators (\ref{eq:com1})--(\ref{eq:com3}) read as
\begin{align}
[J_{-},J_{0}]&=\frac{1}{2}J_{2}+\frac{\alpha_{-}}{2}J_{4},\\
[J_{+},J_{0}]&=\frac{1-\alpha_{+}\alpha_{-}}{2(\alpha_{-})^{2}}z^{3}\left(
 z\frac{\rmd}{\rmd z}J_{1}+\alpha_{-}\frac{\rmd}{\rmd z}J_{4}\right)
 -\frac{1}{2}J_{6}-\frac{\alpha_{+}}{2}J_{7},\\[5pt]
[J_{+},J_{-}]&=-(\alpha_{+}\alpha_{-}+1)J_{3}+(2\alpha_{+}\alpha_{-}-1)
 J_{5}+\alpha_{+}\alpha_{-}.
\end{align}
Thus, they are all second-order operators if and only if $\alpha_{+}
\alpha_{-}=1$. On the other hand, the three operators $J_{i}$ ($i=-,0,+$)
under the condition (\ref{eq:nc1}) have the following form:
\begin{align}
J_{-}=-\alpha_{-}\frac{\rmd}{\rmd z},\quad
 J_{0}=\frac{z}{2}\frac{\rmd}{\rmd z},\quad J_{+}=\frac{\alpha_{+}
 \alpha_{-}-1}{2\alpha_{-}}z^{3}\frac{\rmd^{2}}{\rmd z^{2}}-\frac{
 2\alpha_{+}\alpha_{-}-1}{2\alpha_{-}}z^{2}\frac{\rmd}{\rmd z}+\alpha_{+}z.
\end{align}
Hence, when the condition $\alpha_{+}\alpha_{-}=1$ is satisified, all the
three operators $J_{i}$ ($i=-,0,+$) turn to be of first order and their
Lie algebra is closed simultaneously:
\begin{align}
[J_{-},J_{0}]=\frac{1}{2}J_{-},\qquad [J_{+},J_{0}]=-\frac{1}{2}J_{+},\qquad
 [J_{+},J_{-}]=-2J_{0}+1.
\end{align}
That is exactly what takes place in the Lie-algebraic type A case.
Indeed, each of the operator $J_{i}$ ($i=-,0,+$) is proportional to
the corresponding type A quasi-solvable operator $J_{i}^{(\rmA)}$
($i=-,0,+$) in (\ref{eq:reJsA}) when $\alpha_{+}\alpha_{-}=1$. The above example
demonstrates that a finite set of quasi-solvable operators preserving a given
linear space rarely closes a finite-dimensional Lie algebra. Therefore, we can
assert that quasi-solvable operators which can be constructed by enveloping
algebra of a Lie algebra are rather exceptional.

\section{Type \boldmath{$X_{2}$} 3-fold SUSY with polynomial subspaces}
\label{sec:X2}

In a certain problem, it is more convenient to start with a more general
three-dimensional linear space of functions in a gauged space than
(\ref{eq:tV3-}). Type $X_{2}$ $3$-fold SUSY~\cite{Ta10a} is such an example. 
Hence, we shall first reformulate type B $3$-fold SUSY in a more general 
setting. Instead of the linear
space (\ref{eq:tV3-}), we shall consider as an invariant subspace
\begin{align}
\tcV^{\prime -}_{3}[u]=\braket{\tvph_{1}(u),\tvph_{2}(u),\tvph_{3}(u)},
\label{eq:tV3-'}
\end{align}
in a gauged $u$-space. The relation to the linear space (\ref{eq:tV3-})
in a gauged $z$-space is obviously
\begin{align}
\tcV^{\prime -}_{3}[u]=\tvph_{1}(u)\tcV^{-}_{3}[z]\bigr|_{z=\tvph_{2}(u)
 /\tvph_{1}(u),\,f(z)=\tvph_{3}(u)/\tvph_{1}(u)}\,.
\label{eq:reuz}
\end{align}
With this relation, we can easily produce the most general type B $3$-fold
SUSY in the $u$-space associated with the linear space (\ref{eq:tV3-'}).
Indeed, it follows from (\ref{eq:reuz}) that the gauged type B $3$-fold 
supercharge component $\tP_{3}^{\prime -}[u]$ and the eight linearly
independent second-order linear differential operators $J'_{i}[u]$
($i=1,\dots,8$)
which preserve the linear space (\ref{eq:tV3-'}) in the gauged $u$-space
are constructed from the ones in the $z$-space, namely, $\tP_{3}^{-}[z]$
in (\ref{eq:tP3-}) and $J_{i}[z]$ in (\ref{eq:Js}) via the relations
\begin{align}
\begin{split}
\tP_{3}^{\prime -}[u]&=\tvph_{1}(u)\tP_{3}^{-}[z]\tvph_{1}(u)^{-1}
 \bigl|_{z=\tvph_{2}(u)/\tvph_{1}(u)},\\
 J'_{i}[u]&=\tvph_{1}(u)J_{i}[z]\tvph_{1}(u)^{-1}\bigl|_{z=\tvph_{2}(u)/
 \tvph_{1}(u)}.
\end{split}
\end{align}
Their explicit forms are respectively given by
\begin{align}
\tP_{3}^{\prime -}[u]=u'(q)^{3}\left( \frac{\rmd}{\rmd u}
 +\frac{\tvph'_{1}}{\tvph_{1}}+\frac{W'_{2,1}}{W_{2,1}}
 -\frac{W'_{31,21}}{W_{31,21}}\right)\left( \frac{\rmd}{\rmd u}
 +\frac{\tvph'_{1}}{\tvph_{1}}-\frac{W'_{2,1}}{W_{2,1}}\right)
 \left( \frac{\rmd}{\rmd u}-\frac{\tvph'_{1}}{\tvph_{1}}\right),
\label{eq:tP3-'}
\end{align}
where each of the function $W_{i,j}(u)$ ($W_{ij,kl}(u)$, respectively) is
the Wronskian of $\tvph_{i}(u)$ and $\tvph_{j}(u)$ (of $W_{i,j}(u)$ and 
$W_{k,l}(u)$, respectively),
\begin{align}
W_{i,j}(u)&=\tvph'_{i}(u)\tvph_{j}(u)-\tvph_{i}(u)\tvph'_{j}(u),\\
W_{ij,kl}(u)&=W'_{i,j}(u)W_{k,l}(u)-W_{i,j}(u)W'_{k,l}(u),
\end{align}
and by
\begin{align}
\begin{split}
J'_{1}[u]&=\frac{W_{2,1}(\tvph_{1})^{2}}{W_{31,21}}\left(
 \frac{\rmd^{2}}{\rmd u^{2}}-\frac{W'_{2,1}}{W_{2,1}}\frac{\rmd}{\rmd u}
 +\frac{W'_{2,1}\tvph'_{1}-W_{2,1}\tvph''_{1}}{W_{2,1}\tvph_{1}}\right),\\
J'_{4}[u]&=\frac{W_{3,1}(\tvph_{1})^{2}}{W_{31,21}}\left(
 \frac{\rmd^{2}}{\rmd u^{2}}-\frac{W'_{2,1}}{W_{2,1}}\frac{\rmd}{\rmd u}
 +\frac{W'_{2,1}\tvph'_{1}-W_{2,1}\tvph''_{1}}{W_{2,1}\tvph_{1}}\right)
 -\frac{(\tvph_{1})^{2}}{W_{2,1}}\left(
 \frac{\rmd}{\rmd u}-\frac{\tvph'_{1}}{\tvph_{1}}\right),\\
J'_{2}[u]&=\frac{\tvph_{2}}{\tvph_{1}}J'_{1}[u],\qquad
 J'_{3}[u]=\frac{\tvph_{3}}{\tvph_{1}}J'_{1}[u],\qquad
 J'_{5}[u]=\frac{\tvph_{2}}{\tvph_{1}}J'_{4}[u],\\
J'_{6}[u]&=\frac{\tvph_{3}}{\tvph_{1}}J'_{4}[u],\qquad
 J'_{7}[u]=\frac{\tvph_{2}}{\tvph_{1}}J'_{9}[u],\qquad
 J'_{8}[u]=\frac{\tvph_{3}}{\tvph_{1}}J'_{9}[u],
\end{split}
\label{eq:J's}
\end{align}
where $J'_{9}[u]$ is defined by
\begin{align}
J'_{9}[u]=\frac{W_{3,2}(\tvph_{1})^{2}}{W_{31,21}}\left(
 \frac{\rmd^{2}}{\rmd u^{2}}-\frac{W'_{2,1}}{W_{2,1}}\frac{\rmd}{\rmd u}
 +\frac{W'_{2,1}\tvph'_{1}-W_{2,1}\tvph''_{1}}{W_{2,1}\tvph_{1}}\right)
 -\frac{\tvph_{2}\tvph_{1}}{W_{2,1}}\left( \frac{\rmd}{\rmd u}
 -\frac{\tvph'_{1}}{\tvph_{1}}\right) +1.
\end{align}
The plus component of the gauged type B $3$-fold supercharge
$\bar{P}_{3}^{\prime +}[u]$ and the linear space $\bar{\cV}_{3}^{\prime +}
[u]$ annihilated by it in the $u$-space are related to the ones in the
$z$-space, namely, to $\bar{P}_{3}^{+}[z]$ in (\ref{eq:bP3+}) and to 
$\bar{\cV}_{3}^{+}[z]$ in (\ref{eq:bV3+}), respectively, as
\begin{align}
\begin{split}
\bar{P}_{3}^{\prime +}[u]&=\tvph_{1}(u)^{3}W_{2,1}(u)^{-2}\bar{P}_{3}^{+}[z]
 W_{2,1}(u)^{2}\tvph_{1}(u)^{-3}\bigl|_{z=\tvph_{2}(u)/\tvph_{1}(u)}\,,\\
\bar{\cV}_{3}^{\prime +}[u]&=\tvph_{1}(u)^{3}
 W_{2,1}(u)^{-2}\bar{\cV}_{3}^{+}[z]\bigl|_{z=\tvph_{2}(u)/\tvph_{1}(u)}\,.
\end{split}
\end{align}
Their explicit forms are respectively given by
\begin{align}
\bar{P}_{3}^{\prime +}[u]=-u'(q)^{3}\left(\frac{\rmd}{\rmd u}
 +\frac{\tvph'_{1}}{\tvph_{1}}\right)\left(\frac{\rmd}{\rmd u}
 -\frac{\tvph'_{1}}{\tvph_{1}}+\frac{W'_{2,1}}{W_{2,1}}\right)
 \left(\frac{\rmd}{\rmd u}-\frac{\tvph'_{1}}{\tvph_{1}}
 -\frac{W'_{2,1}}{W_{2,1}}+\frac{W'_{31,21}}{W_{31,21}}\right),
\label{eq:bP3+'}
\end{align}
and by
\begin{align}
\bar{\cV}_{3}^{\prime +}[u]=\frac{\tvph_{1}(u)}{W_{31,21}(u)}
 \braket{W_{2,1}(u), W_{3,1}(u), W_{3,2}(u)}.
\label{eq:bV3+'}
\end{align}
The eight linearly independent second-order linear differential operators
$K'_{i}[u]$ ($i=1,\dots,8$) which preserve the latter space (\ref{eq:bV3+'})
are also constructed from $K_{i}[z]$ in (\ref{eq:Ks}) via the relation
\begin{align}
K'_{i}[u]=\tvph_{1}(u)^{3}W_{2,1}(u)^{-2}K_{i}[z]W_{2,1}(u)^{2}\tvph_{1}
 (u)^{-3}\bigl|_{z=\tvph_{2}(u)/\tvph_{1}(u)}\,,
\end{align}
and are explicitly given by
\begin{align}
\begin{split}
K'_{1}[u]=&\;\frac{W_{2,1}(\tvph_{1})^{2}}{W_{31,21}}\biggl[
 \frac{\rmd^{2}}{\rmd u^{2}}-\left( \frac{2\tvph'_{1}}{\tvph_{1}}
 -\frac{W'_{31,21}}{W_{31,21}}\right)\frac{\rmd}{\rmd u}
 -\frac{\tvph''_{1}}{\tvph_{1}}-\frac{W''_{2,1}}{W_{2,1}}\\
&\;+\frac{W''_{31,21}}{W_{31,21}}+\frac{2(\tvph'_{1})^{2}}{(\tvph_{1})^{2}}
 -\left(\frac{\tvph'_{1}}{\tvph_{1}}-\frac{W'_{2,1}}{W_{2,1}}
 +\frac{W'_{31,21}}{W_{31,21}}\right)\frac{W'_{31,21}}{W_{31,21}}\biggr],\\
K'_{2}[u]=&\;\frac{\tvph_{2}}{\tvph_{1}}K'_{1}[u]-K'_{0}[u],\quad K'_{3}[u]=
 \frac{\tvph_{3}}{\tvph_{1}}K'_{1}[u]-\frac{W_{3,1}}{W_{2,1}}K'_{0}[u]+1,\\
K'_{4}[u]=&\;\frac{W_{3,1}}{W_{2,1}}K'_{1}[u],\quad K'_{5}[u]=
 \frac{W_{3,1}}{W_{2,1}}K'_{2}[u],\quad
 K'_{6}[u]=\frac{W_{3,1}}{W_{2,1}}K'_{3}[u],\\
K'_{7}[u]=&\;\frac{W_{3,2}}{W_{2,1}}K'_{2}[u],\quad
 K'_{8}[u]=\frac{W_{3,2}}{W_{2,1}}K'_{3}[u],
\end{split}
\label{eq:K's}
\end{align}
where $K'_{0}$ is defined by
\begin{align}
K'_{0}[u]=\frac{(W_{2,1})^{2}}{W_{31,21}}\left( \frac{\rmd}{\rmd u}
 -\frac{\tvph'_{1}}{\tvph_{1}}-\frac{W'_{2,1}}{W_{2,1}}
 +\frac{W'_{31,21}}{W_{31,21}}\right).
\end{align}
With these preliminaries, let us construct the most general type $X_{2}$
$3$-fold SUSY system. The three-dimensional type $X_{2}$ polynomial
subspace is introduced as (\ref{eq:tV3-'}) with $\tvph_{n}(u)$ ($n=1,2,3$)
given by~\cite{Ta10a}
\begin{align}
\tvph_{n}(u;\alpha)=(\alpha+n-2) u^{n+1}+2(\alpha+n-1)(\alpha-1) u^{n}
 +(\alpha+n)(\alpha-1) \alpha u^{n-1},
\label{eq:3X2}
\end{align}
where $\alpha$ is a free parameter. The gauged type $X_{2}$ $3$-fold
supercharge component is calculated by using the formula (\ref{eq:tP3-'}) as
\begin{align}
\tP_{3}^{\prime -}[u;\alpha]=u'(q)^{3}\frac{f_{\alpha}}{f_{\alpha+2}}\left(
 \frac{\rmd}{\rmd u}-\frac{f'_{\alpha+3}}{f_{\alpha+3}}\right)
 \frac{f_{\alpha+2}}{f_{\alpha+1}}\left( \frac{\rmd}{\rmd u}
 -\frac{f'_{\alpha+2}}{f_{\alpha+2}}\right)\frac{f_{\alpha+1}}{f_{\alpha}}
 \left(\frac{\rmd}{\rmd u}-\frac{f'_{\alpha+1}}{f_{\alpha+1}}\right),
\end{align}
where the function $f_{\alpha}(u)$ is given by
\begin{align}
f_{\alpha}(u)=f(u;\alpha)=u^{2}+2(\alpha-1) u+(\alpha-1)\alpha.
\end{align}
It exactly coincides with Eq.~(3.6) for $\cN=3$ in Ref.~\cite{Ta10a}.
However, the latter reference reported only four linearly independent
second-order linear differential operators $J_{i}^{(X_{2})}(\alpha)$
($i=1,\dots,4$), summarized in (\ref{eq:ApJsX2a}) and (\ref{eq:ApJsX2b}),
which left the type $X_{2}$ polynomial subspace (\ref{eq:tV3-'}) with
(\ref{eq:3X2}) invariant. A direct calculation of the eight
operators $J'_{i}[u]$ in (\ref{eq:J's}) using (\ref{eq:3X2}) shows that
the former are expressible as linear combinations of the latter as
\begin{align}
J_{i}^{(X_{2})}(\alpha)=\sum_{j=1}^{8}C_{ij}(\alpha)J'_{j}[u;\alpha]
 +C_{i0}(\alpha)\qquad (i=1,\dots,4),
\label{eq:JsX2}
\end{align}
where $C_{ij}(\alpha)$ ($i=1,\dots,4;\,j=0,\dots,8$) are all constants
whose explicit forms are given in Appendix~\ref{app:coef}.
Hence, there are four other linear combinations of $J'_{i}[u;\alpha]$
($i=1,\dots,8$) which
are linearly independent of $J_{i}^{(X_{2})}(\alpha)$ ($i=1,\dots,4$).

The plus component of the gauged type $X_{2}$ $3$-fold supercharge
$\bar{P}_{3}^{\prime +}$ is similarly calculated from the formula
(\ref{eq:bP3+'}) with (\ref{eq:3X2}) as
\begin{align}
\bar{P}_{3}^{\prime +}[u]=u'(q)^{3}\left(\frac{\rmd}{\rmd u}
 +\frac{f'_{\alpha+1}}{f_{\alpha+1}}\right)\frac{f_{\alpha+1}}{f_{\alpha}}
 \left(\frac{\rmd}{\rmd u}+\frac{f'_{\alpha+2}}{f_{\alpha+2}}\right)
 \frac{f_{\alpha+2}}{f_{\alpha+1}}\left(\frac{\rmd}{\rmd u}
 +\frac{f'_{\alpha+3}}{f_{\alpha+3}}\right)\frac{f_{\alpha}}{f_{\alpha+2}},
\end{align}
which exactly coincides with Eq.~(3.18) for $\cN=3$ in Ref.~\cite{Ta10a}.
The linear space $\bar{\cV}_{3}^{\prime +}[u]$ (\ref{eq:bV3+'}) annihilated
by it reads by using (\ref{eq:3X2}) as
\begin{align}
\bar{\cV}_{3}^{\prime +}[u;\alpha+3]&=f_{\alpha}(u)^{-1}f_{\alpha+3}(u)^{-1}
 \braket{\bar{\chi}_{1}(u,\alpha+3), \bar{\chi}_{2}(u,\alpha+3),
 \bar{\chi}_{3}(u,\alpha+3)}\notag\\
&=f_{\alpha}(u)^{-1}f_{\alpha+3}(u)^{-1}\tcV_{3}^{(X_{2b})}[u;\alpha+3],
\label{eq:X2b}
\end{align}
where the function $\bar{\chi}_{n}(u;\alpha)$ ($n=1,2,3$) is defined by
\begin{align}
\bar{\chi}_{n}(u;\alpha)=&\;(\alpha-n)(\alpha-n+1) u^{n+1}+2(\alpha-n-1)
 (\alpha-n+1)(\alpha-1) u^{n}\notag\\
&\;+(\alpha-n-1)(\alpha-n)(\alpha-1)\alpha u^{n-1},
\label{eq:3X3}
\end{align}
and thus reproduces Eq.~(3.20) for $\cN=3$ in Ref.~\cite{Ta10a}. It is
evident that there are eight linearly independent second-order linear 
differential operators $\tK'_{i}[u;\alpha+3]$ which preserve the polynomial 
subspace $\tcV_{3}^{(X_{2b})}[u;\alpha+3]$ appeared in (\ref{eq:X2b}).
Since  $K'_{i}[u]$ ($i=1,\dots,8$) in (\ref{eq:K's}) leave the linear space
$\bar{\cV}_{3}^{\prime +}[u;\alpha+3]$
invariant, it follows from (\ref{eq:X2b}) that they are given by
\begin{align}
\tK'_{i}[u;\alpha+3]=f_{\alpha}(u)^{-1}f_{\alpha+3}(u)^{-1}
 K'_{i}[u;\alpha+3] f_{\alpha+3}(u)f_{\alpha}(u)\quad (i=1,\dots,8).
\label{eq:tKs}
\end{align}
There were again only four linearly independent second-order linear
differential operators $K_{i}^{(X_{2})}(\alpha)$ in Ref.~\cite{Ta10a}, 
summarized in (\ref{eq:ApKsX2a}) and (\ref{eq:ApKsX2b}), which
preserved the linear space $\tcV_{3}^{(X_{2b})}[u;\alpha]$.
A direct calculation of the eight operators $\tK'_{i}[u;\alpha+3]$
in (\ref{eq:tKs}) using (\ref{eq:K's}) and (\ref{eq:3X2}) shows that
the former are expressible as linear combinations of the latter as
\begin{align}
K_{i}^{(X_{2})}(\alpha)=\sum_{j=1}^{8}C_{ij}(\alpha-3)\tK'_{j}[u;\alpha]
 +C_{i0}(\alpha-3)\qquad (i=1,\dots,4),
\label{eq:KsX2}
\end{align}
where $C_{ij}(\alpha-3)$ ($i=1,\dots,4;\,j=0,\dots,8$) are all constants
whose explicit forms are also given in Appendix~\ref{app:coef}.
Hence, there are four other linear combinations of $\tK'_{i}[u;\alpha]$
($i=1,\dots,8$) which
are linearly independent of $K_{i}^{(X_{2})}(\alpha)$ ($i=1,\dots,4$).

\section{Discussion and Summary}
\label{sec:discus}

In this paper, we have shown that type B $3$-fold SUSY is a necessary
and sufficient condition for a Schr\"{o}dinger operator to admit
three linearly independent local analytic solutions, and thus to preserve
a three-dimensional non-polynomial linear space. The most general
operator of this kind consists of eight linearly independent non-trivial
differential operators of at most second order. As a by-product of
the latter finding, we have found that all the known $3$-fold SUSY
associated with monomial or polynomial subspaces, which are regarded
as particular cases of the most general type B $3$-fold SUSY,
also admit eight operators for each component Hamiltonian,
some of which have been missed in the literature.
In particular, the three different types of $3$-fold SUSY associated
with monomial subspaces, namely, type A, type B, and type C, are
connected continuously via a parameter.
The results obtained here are full of implications for the future
studies, and we would like to close this paper by referring to some of
them.\\

The fact that the number of linearly independent quasi-solvable operators
of at most second order preserving a three-dimensional linear function
space is always the same casts a natural question: Is it true for any
dimensional linear function space? According to the classification of
quasi-solvable operators which leave a monomial space invariant done in
Ref.~\cite{PT95}, there are no extra operators in the cases of type B and
C monomial subspaces (the general case (4) and the special case C and D
there) except for the three-dimensional and a particular four-dimensional
ones. The number of linearly independent non-trivial quasi-solvable
operators of at most second order preserving a type A monomial subspace
(the general case (3) in Ref.~\cite{PT95}), on the other hand,
is always eight regardless of the dimension of the subspace.
A mathematical theorem (Corollary 2) in Ref.~\cite{GKM05}
assures the latter, and further states that the number of
linearly independent at most second-order linear differential operators
preserving a type B monomial space of dimension greater than three is,
excluding the trivial constant multiplication operation, six.
Hence, it does not seem to be true in the cases of function spaces of
more than three dimension.

Let us next observe the problem from a different point of view. A crucial
consequence of the fact in the present three-dimensional case is that
the eight linearly independent quasi-solvable operators preserving monomial
spaces of type A, type B, and type C are connected continuously with each
other via one parameter. In this respect, we note that these three
different types are connected continuously, at least at the level of
monomial subspaces, also in any dimension of more than three. In fact,
an $\cN$-dimensional type C monomial subspace (\ref{eq:typeC}) with
$\cN_{1}=\cN-1$ and $\cN_{2}=1$ reads as
\begin{align}
\tcV_{\cN-1,\,1}^{(\rmC)}=\braket{1,z,\dots,z^{\cN-2},z^{\lambda}}.
\end{align}
Evidently, it reduces to an $\cN$-dimensional type B monomial subspace
when $\lambda=\cN$ and to a type A one when $\lambda=\cN-1$. Hence, in
contrast with the previous observation, it seems that the latter two types
are connected continuously with the particular case of type C via the
parameter $\lambda$ also in any dimension $\cN>3$.

If it is indeed the case, the number of linearly independent
quasi-solvable operators in the cases of type B and type C must be the same
as the one in the type A case, namely, eight. If it is not the case, on
the other hand, some irregular phenomena would take place in the limits
$\lambda\to 1$ and $2$. It could be superficial, however, arising from
the restriction to at most second-order linear differential operators.
Actually, if we restrict our consideration to first-order ones in the present
$\cN=3$ case, they are no longer connected continuously with each other and
the number of linearly independent operators varies (one in type B
and type C, three in type A, and zero in general non-monomial cases).
Hence, it could be the case that the number is invariable for operators
of a certain order higher than two in a linear
function space of dimension higher than three.

Another intriguing fact is that there is at least one extra quasi-solvable 
second-order operator for a particular four-dimensional type C
monomial subspace with $\cN_{1}=\cN_{2}=2$ (the special case B in
Ref.~\cite{PT95}):
\begin{align}
\tcV_{2,2}^{(\rmC)}=\braket{1,z,z^{\lambda},z^{\lambda+1}}.
\end{align}
Hence, we expect that a detailed examination for the four-dimensional
case would give us further insights into the issue.\\

The fact that the most general type B $3$-fold SUSY system admits three
linearly independent local solutions of non-polynomial type in closed form
indicates that there would be a possibly tremendous number of Schr\"{o}dinger
operators having non-polynomial local solutions which have not been
found yet. Until now, virtually all the Schr\"{o}dinger operators known to
have local analytic solutions and thus be (quasi-)solvable are those whose 
solutions are expressible as products of a polynomial in a particular
function and a gauge factor (the ground-state wave function when exactly 
solvable).

As is well-known, $\cN$-fold SUSY is equivalent to weak quasi-solvability
for one-dimensional Schr\"{o}dinger operators and hence any of them admitting
local solutions in closed form possesses (at least) one type of $\cN$-fold
SUSY. There have been four types so far found, and all the constructed
$\cN$-fold SUSY systems are those having polynomial-type local solutions,
or equivalently, those preserving a polynomial subspace. It should be noted,
however, that except for type A they constitute only subclasses of the most
general $\cN$-fold SUSY in each type. In the case of type A, it was already
proved \cite{ANST01,Ta03a} that it is equivalent to the $\fsl(2)$
Lie-algebraic quasi-solvable models \cite{Tu88} preserving a type A
monomial space.
On the other hand, all the other types of $\cN$-fold SUSY systems were
constructed under a particular restriction $F(q)=z'(q)/z(q)$ which
guarantees that a linear function space preserved by the operator under
consideration is of monomial or polynomial type. In other words, linear
differential operators admitting an invariant subspace of non-polynomial
type have not been systematically considered yet.

To explore the area of operators with a non-polynomial invariant subspace,
it is now evident that we must tackle a necessary and sufficient condition
for other types of $\cN$-fold SUSY, or at least examine them under a less
restrictive condition than $F(q)=z'(q)/z(q)$. The present investigation on
the necessary and sufficient condition for type B $3$-fold SUSY would
definitely provide us the first clue and foothold toward that research
direction.\\

The notion of shape invariance introduced in Ref.~\cite{Ge83} is
a well-known tool for investigating and constructing an exactly solvable
Schr\"{o}dinger operator. Precisely speaking, it is a sufficient
condition for \emph{solvability} but is neither necessary nor sufficient
for \emph{exact solvability}. Later in Ref.~\cite{BDGKPS93}, it was
generalized to the notion of two- and multi-step shape invariance,
keeping the  sufficiency for solvability intact. Recently,
it was shown \cite{RT13} that any two-step shape-invariant model can be
systematically investigated and constructed in the framework of $\cN$-fold
SUSY as a particular case of $2$-fold SUSY with an intermediate Hamiltonian
developed in Ref.~\cite{BT09}. Now that we appreciate the most general
$3$-fold SUSY which is quasi-solvable \emph{in the strong sense}, we are
in a position to
follow the latter program to uncover the area of three-step shape invariance.
In the case of type A $3$-fold SUSY, it was already shown \cite{BT10} that
the possible three-step shape-invariant potentials are all \emph{reducible},
that is, they have ordinal shape invariance as well. Hence, we would be able
to see whether there exists an \emph{irreducible} three-step shape-invariant
potential by using the results here.



\appendix

\section{Quasi-solvable operators in type C $3$-fold SUSY}
\label{app:C}

In Ref.~\cite{PT95}, the set of quasi-solvable operators up to second order 
preserving the type C monomial subspace (\ref{eq:typeC}) was first 
investigated and four were found for an arbitrary $\cN_{1},\cN_{2}\in\bbN$,
cf.\ the general case (4-a) there. It was also reexamined in Proposition~3,
Ref.~\cite{GKM05}, which would be applicable only for $\cN_{1}+\cN_{2}>3$,
though it was not stated explicitly there, assumed in the preceding Theorem~2.
In our present case of $\cN_{1}=2$ and $\cN_{2}=1$, they reduce to
(cf.\ (3.7a)--(3.7d) in Ref.~\cite{GT05})
\begin{align}
\begin{split}
&J_{0}^{(\rmC)}(\lambda)=z\frac{\rmd}{\rmd z},\qquad J_{0-}^{(\rmC)}
 (\lambda)=\frac{\rmd}{\rmd z}\left( z\frac{\rmd}{\rmd z}-\lambda\right),\\
&J_{00}^{(\rmC)}(\lambda)=z^{2}\frac{\rmd^{2}}{\rmd z^{2}},\qquad
 J_{+0}^{(\rmC)}(\lambda)=z\left( z\frac{\rmd}{\rmd z}-1\right)\left(
 z\frac{\rmd}{\rmd z}-\lambda\right).
\end{split}
\label{eq:ApJsC}
\end{align}
In addition to the above, there exist two extra second-order operators
preserving the three-dimensional type C monomial space (\ref{eq:tV3-C}).
This is because three-dimensional monomial subspaces are special and
it can also be regarded as the special case A (a) in Ref.~\cite{PT95}.
The two additional operators are given by
\begin{align}
J_{\# -}^{(\rmC)}(\lambda)=z^{\lambda}\frac{\rmd}{\rmd z}\left(
 z\frac{\rmd}{\rmd z}-\lambda\right),\qquad
 J_{\# 0}^{(\rmC)}(\lambda)=z^{\lambda}\left( z\frac{\rmd}{\rmd z}
 -1\right)\left( z\frac{\rmd}{\rmd z}-\lambda\right),
\label{eq:ApJsC'}
\end{align}
and were not considered in Ref.~\cite{GT05}.

The other set of quasi-solvable operators related with the above via
$3$-fold SUSY preserves the linear space $\bar{\cV}_{3}^{+}$ in
(\ref{eq:bV3+C}) which is equivalent up to a similarity transformation
to another type C monomial subspace $\langle 1,z,z^{1-\lambda}\rangle$.
Hence, it is evident that they are given by $K_{0}^{(\rmC)}(\lambda)=
z\,J_{0}^{(\rmC)}(1-\lambda)\,z^{-1}$ and $K_{ij}^{(\rmC)}(\lambda)=z\,
J_{ij}^{(\rmC)}(1-\lambda)\,z^{-1}$ ($i,j=+,0,-$), cf.\ (3.29) in
Ref.~\cite{GT05}. Explicitly, their forms read as
\begin{align}
\begin{split}
&K_{0}^{(\rmC)}(\lambda)=z\frac{\rmd}{\rmd z}-1,\qquad K_{0-}^{(\rmC)}
 (\lambda)=z^{-1}\left( z\frac{\rmd}{\rmd z}-1\right)\left(
 z\frac{\rmd}{\rmd z}-2+\lambda\right),\\
&K_{00}^{(\rmC)}(\lambda)=\left(z\frac{\rmd}{\rmd z}-2\right)\left(
 z\frac{\rmd}{\rmd z}-1\right),\quad
 K_{+0}^{(\rmC)}(\lambda)=z\left( z\frac{\rmd}{\rmd z}-2\right)\left(
 z\frac{\rmd}{\rmd z}-2+\lambda\right),\\
&K_{\# -}^{(\rmC)}(\lambda)=z^{-\lambda}\left( z\frac{\rmd}{\rmd z}
 -1\right)\left( z\frac{\rmd}{\rmd z}-2+\lambda\right),\\
&K_{\# 0}^{(\rmC)}(\lambda)=z^{1-\lambda}\left( z\frac{\rmd}{\rmd z}
 -2\right)\left( z\frac{\rmd}{\rmd z}-2+\lambda\right).
\end{split}
\label{eq:ApKsC}
\end{align}
The last two operators were not considered in Ref.~\cite{GT05}.

\section{Quasi-solvable operators in type B $3$-fold SUSY}
\label{app:B}

The set of quasi-solvable operators up to second order preserving the
type B monomial subspace (\ref{eq:typeB}) was also investigated in
Ref.~\cite{PT95} and six were found for an arbitrary $\cN\in\bbN$.
It was also restudied in Corollary~2 of Ref.~\cite{GKM05}, which would
be applicable only for $\cN>3$ assumed in the preceding Theorem 2.
In the case of $\cN=3$, they reduce to (cf.\ (3.11a)--(3.11f) in
Ref.~\cite{GT04})
\begin{align}
\begin{split}
&J_{0}^{(\rmB)}=z\frac{\rmd}{\rmd z},\quad J_{--}^{(\rmB)}=\frac{\rmd^{2}
 }{\rmd z^{2}},\quad J_{0-}^{(\rmB)}=\frac{\rmd}{\rmd z}\left(
 z\frac{\rmd}{\rmd z}-3\right),\quad
 J_{00}^{(\rmB)}=z^{2}\frac{\rmd^{2}}{\rmd z^{2}},\\
&J_{+0}^{(\rmB)}=z\left( z\frac{\rmd}{\rmd z}-3\right)\left(
 z\frac{\rmd}{\rmd z}-1\right),\qquad J_{++}^{(\rmB)}=z^{3}
 \frac{\rmd}{\rmd z}\left( z\frac{\rmd}{\rmd z}-3\right).
\end{split}
\label{eq:ApJsB}
\end{align}
In addition to the above, there exists an extra second-order operator
preserving the three-dimensional type B monomial space
(cf.\ the special case A (b) in Ref.~\cite{PT95}):
\begin{align}
J_{3+}^{(\rmB)}=z^{3}\left(z\frac{\rmd}{\rmd z}-3\right)\left(
 z\frac{\rmd}{\rmd z}-1\right).
\label{eq:ApJsB'}
\end{align}
The latter operator was not considered in Ref.~\cite{GT04}.

The other set of quasi-solvable operators related with the above via
$3$-fold SUSY, which preserves the linear space $\bar{\cV}_{3}^{+}$ in
(\ref{eq:bV3+B}), reads as (cf.\ (3.16a)--(3.16f) in Ref.~\cite{GT04})
\begin{align}
\begin{split}
&K_{0}^{(\rmB)}=z\frac{\rmd}{\rmd z},\qquad K_{--}^{(\rmB)}=z^{-2}\left(
 z\frac{\rmd}{\rmd z}+1\right)\left( z\frac{\rmd}{\rmd z}-2\right),\\
&K_{0-}^{(\rmB)}=z^{-1}\left( z\frac{\rmd}{\rmd z}+1\right)\left(
 z\frac{\rmd}{\rmd z}
 -1\right),\qquad K_{00}^{(\rmB)}=z^{2}\frac{\rmd^{2}}{\rmd z^{2}},\\
&K_{+0}^{(\rmB)}=z\left( z\frac{\rmd}{\rmd z}-2\right)\left(
 z\frac{\rmd}{\rmd z}+1\right),\qquad K_{++}^{(\rmB)}=z^{2}\left(
 z\frac{\rmd}{\rmd z}-2\right)\left(z\frac{\rmd}{\rmd z}-1\right).
\end{split}
\label{eq:ApKsB}
\end{align}
The space $\bar{\cV}_{3}^{+}$ is equivalent to a three-dimensional
monomial subspace $\langle 1,z^{2},z^{3}\rangle$ up to a similarity
transformation. Hence, the latter monomial space is preserved by
$z\,K_{0}^{(\rmB)}\,z^{-1}$ and $z\,K_{ij}^{(\rmB)}\,z^{-1}$ ($i,j=+,0,-$).
A set of quasi-solvable operators of up to second order
preserving it was also examined in Ref.~\cite{PT95} as the special
case A (c) with $m=2$ and six were presented. Transformed back to the space 
$\bar{\cV}_{3}^{+}$ in (\ref{eq:bV3+B}), they correspond to $K_{0}^{(\rmB)}$,
$K_{0-}^{(\rmB)}$, $K_{00}^{(\rmB)}$, $K_{+0}^{(\rmB)}$, $K_{++}^{(\rmB)}$,
and another one given by
\begin{align}
K_{3+}^{(\rmB)}=z^{3}\left(z\frac{\rmd}{\rmd z}-2\right)\left(
 z\frac{\rmd}{\rmd z}-1\right).
\label{eq:ApKsB'}
\end{align}
Hence, there have been in total seven linearly independent ones so far
found in Refs.~\cite{PT95,GT04}.

\section{Quasi-solvable operators in type A $3$-fold SUSY}
\label{app:A}

The set of quasi-solvable operators up to second order preserving
the type A monomial subspace (\ref{eq:typeA}) is exhausted by the
first-order differential operator representation of $\fsl(2)$ Lie algebra
and their quadratic form, which was first reported in Ref.~\cite{Tu88}
(cf.\ (4.13) and (4.17)--(4.21) in Ref.~\cite{ANST01}):
\begin{align}
\begin{split}
&J_{-}^{(\rmA)}=\frac{\rmd}{\rmd z},\qquad J_{0}^{(\rmA)}=z\frac{\rmd}{
 \rmd z},\qquad J_{+}^{(\rmA)}=z\left( z\frac{\rmd}{\rmd z}-2\right),\\
&J_{--}^{(\rmA)}=\frac{\rmd^{2}}{\rmd z^{2}},\qquad J_{0-}^{(\rmA)}=
 z\frac{\rmd^{2}}{\rmd z^{2}},\qquad J_{00}^{(\rmA)}=z^{2}
 \frac{\rmd^{2}}{\rmd z^{2}},\\
&J_{+0}^{(\rmA)}=z^{2}\frac{\rmd}{\rmd z}\left( z\frac{\rmd}{\rmd z}
 -2\right),\qquad J_{++}^{(\rmA)}=z^{2}\left( z\frac{\rmd}{\rmd z}
 -2\right)\left( z\frac{\rmd}{\rmd z}-1\right).
\end{split}
\label{eq:ApJsA}
\end{align}
The other set of quasi-solvable operators related with the above via
$3$-fold SUSY preserves the same type A monomial subspace (\ref{eq:typeA})
and thus is the same, as shown in Ref.~\cite{Ta03a} for arbitrary
$\cN\in\bbN$. Hence, we have $K_{i}^{(\rmA)}=J_{i}^{(\rmA)}$ and
$K_{ij}^{(\rmA)}=J_{ij}^{(\rmA)}$ ($i=+,0,-$).

\section{Quasi-solvable operators in type $X_{2}$ $3$-fold SUSY}
\label{app:X2}

In Ref.~\cite{Ta10a}, second-order quasi-solvable operators preserving
a type $X_{2}$ exceptional polynomial subspace were first studied and
four linearly independent ones were found for each subspace. In the case
of three-dimensional subspace $\tcV^{-'}_{3}[u;\alpha]=\tcV^{(X_{2a})}_{3}
[u;\alpha]$, namely, (\ref{eq:tV3-'}) with (\ref{eq:3X2}), they reduce to
(cf.\ (2.12), (2.15), (2.18), and (2.20) in Ref.~\cite{Ta10a})
\begin{align}
\begin{split}
J_{1}^{(X_{2})}(\alpha)=&\;u\frac{\rmd^{2}}{\rmd u^{2}}-(u-\alpha+3)
 \frac{\rmd}{\rmd u}+\frac{4(\alpha-1)(u+\alpha)}{f(u;\alpha)}\left(
 \frac{\rmd}{\rmd u}-1\right),\\
J_{2}^{(X_{2})}(\alpha)=&\;\left[ u^{2}+(\alpha+2)(\alpha-1)\right]
 \frac{\rmd^{2}}{\rmd u^{2}}
 -\left[ u^{2}+4 u+(\alpha-1)(3\alpha+2)\right]\frac{\rmd}{\rmd u}+4 u\\
&\;-8(\alpha-1)\frac{\alpha u+\alpha^{2}-1}{f(u;\alpha)}\left(
 \frac{\rmd}{\rmd u}-1\right),\\
J_{3}^{(X_{2})}(\alpha)=&\;(u+2\alpha+2)u^{2}\frac{\rmd^{2}}{\rmd u^{2}}
 +\left[(\alpha-5)u^{2}+(3\alpha^{2}+\alpha-8)u+4(\alpha+4)(\alpha-1)
 \right]\frac{\rmd}{\rmd u}\\
&\;-4(\alpha-2)u-4(\alpha-1)\frac{(\alpha^{2}+3\alpha-8)u+(\alpha+4)
 (\alpha-1)\alpha}{f(u;\alpha)}\left(\frac{\rmd}{\rmd u}-1\right),
\end{split}
\label{eq:ApJsX2a}
\end{align}
and
\begin{align}
\lefteqn{
2(\alpha+1)J_{4}^{(X_{2})}(\alpha)=\left[2(\alpha+1)u+3\alpha^{2}
 +7\alpha+6\right]u^{3}\frac{\rmd^{2}}{\rmd u^{2}}-\bigl[
 12(\alpha+1)u^{3}}\hspace{30pt}\notag\\[5pt]
&+(3\alpha^{3}+8\alpha^{2}+15\alpha+22) u^{2}+(\alpha-1)(7\alpha^{3}
 +27\alpha^{2}+10\alpha-16)u\notag\\[5pt]
&+2(\alpha-1)(\alpha^{4}+8\alpha^{3}+29\alpha^{2}-6\alpha-40)\bigr]
 \frac{\rmd}{\rmd u}\notag\\[5pt]
&+24(\alpha+1) u^{2}+4(\alpha-1)(3\alpha^{2}+2\alpha-4) u\notag\\[5pt]
&+4(\alpha-1)^{2}\frac{(\alpha^{3}+9\alpha^{2}-22\alpha-40) u
 +(\alpha^{3}+9\alpha^{2}-6\alpha-20)\alpha}{f(u;\alpha)}\left(
 \frac{\rmd}{\rmd u}-1\right).
\label{eq:ApJsX2b}
\end{align}
The other set of quasi-solvable operators related with the above via
$3$-fold SUSY, which preserves the other $X_{2}$ exceptional polynomial
subspace $\tcV_{3}^{(X_{2b})}[u;\alpha]$ defined in (\ref{eq:X2b}) with
(\ref{eq:3X3}), reads as
(cf.\ (3.38), (3.41), (3.44), and (3.46) in Ref.~\cite{Ta10a})
\begin{align}
\begin{split}
K_{1}^{(X_{2})}(\alpha)=&\;u\frac{\rmd^{2}}{\rmd u^{2}}+(u-\alpha-3)
 \frac{\rmd}{\rmd u}+\frac{4}{f(u;\alpha)}\left[ (\alpha-1)(u+\alpha)
 \frac{\rmd}{\rmd u}+\alpha(u+\alpha-1)\right],\\
K_{2}^{(X_{2})}(\alpha)=&\;\left[ u^{2}+(\alpha-1)(\alpha-4)\right]
 \frac{\rmd^{2}}{\rmd u^{2}}+\left[ u^{2}-6 u+(\alpha-1)(3\alpha-8)\right]
 \frac{\rmd}{\rmd u}-4 u\\
&\;-\frac{8(\alpha-1)}{f(u;\alpha)}\left\{ [(\alpha-3) u+(\alpha-1)(\alpha-2)]
 \frac{\rmd}{\rmd u}+(\alpha-2) u+\alpha^{2}-3\alpha+4\right\},\\
K_{3}^{(X_{2})}(\alpha)=&\;(u+2\alpha-4) u^{2}\frac{\rmd^{2}}{\rmd u^{2}}
 -\left[(\alpha+3) u^{2}+(3\alpha^{2}-5\alpha-4) u-4(\alpha-1)(\alpha-2)
 \right]\frac{\rmd}{\rmd u}\\
&\;+4\alpha u-\frac{4(\alpha-1)}{f(u;\alpha)}\biggl\{ \left[(\alpha^{2}
 -3\alpha+4) u+\alpha(\alpha-1)(\alpha-2)\right]\frac{\rmd}{\rmd u}\\
&\;+\alpha[(\alpha-2) u+\alpha(\alpha-3)]\biggr\},
\end{split}
\label{eq:ApKsX2a}
\end{align}
and
\begin{align}
\lefteqn{
2(\alpha-2)K_{4}^{(X_{2})}(\alpha)=\left[ 2(\alpha-2) u+3\alpha^{2}-11\alpha
 +12\right] u^{3}\frac{\rmd^{2}}{\rmd u^{2}}-\bigl[ 12(\alpha-2) u^{3}
}\hspace{30pt}\notag\\[5pt]
&-( 3\alpha^{3}-36\alpha^{2}+97\alpha-84) u^{2}-(\alpha-1)( 7\alpha^{3}
 -49\alpha^{2}+112\alpha-80) u\notag\\[5pt]
&-2(\alpha-1)(\alpha^{4}-11\alpha^{3}+32\alpha^{2}-36\alpha+16)\bigr]
 \frac{\rmd}{\rmd u}+24(\alpha-2) u^{2}\notag\\
&-4(3\alpha^{3}-27\alpha^{2}+64\alpha-48) u+\frac{4(\alpha-1)}{f(u;\alpha)}
 \biggl\{ (\alpha-1)\bigl[ (\alpha^{3}-3\alpha^{2}+8\alpha-16) u\notag\\
&+\alpha(\alpha^{3}-3\alpha^{2}+8\alpha-16)\bigr]\frac{\rmd}{\rmd u}
 +\alpha\bigl[ (\alpha^{3}-3\alpha^{2}+8\alpha-16) u\notag\\
&+\alpha(\alpha-1)(\alpha^{2}-3\alpha+4)\bigr]\biggr\},
\label{eq:ApKsX2b}
\end{align}

\section{Commutation Relations of $J_{i}$}
\label{app:CR}

Every commutation relation of the quasi-solvable operators $J_{i}$
($i=1,\dots,8$) is presented in terms of $J_{1}$, $J_{4}$, and $J_{9}$:
\begin{align*}
&{}[J_{1},J_{2}]=\frac{2}{f''}\frac{\rmd}{\rmd z}J_{1},\quad
 [J_{1},J_{3}]=\left(\frac{2f'}{f''}\frac{\rmd}{\rmd z}+1\right)
 J_{1},\quad [J_{1},J_{4}]=2\frac{\rmd}{\rmd z}J_{1},\\
&{}[J_{1},J_{5}]=2z\frac{\rmd}{\rmd z}J_{1}+\frac{2}{f''}\frac{\rmd}{\rmd z}
 J_{4},\quad [J_{1},J_{6}]=2f\frac{\rmd}{\rmd z}J_{1}+\left(
 \frac{2f'}{f''}\frac{\rmd}{\rmd z}+1\right)J_{4},\\
&{}[J_{1},J_{7}]=z\left(2z\frac{\rmd}{\rmd z}-1\right)J_{1}+\frac{2}{f''}
 \frac{\rmd}{\rmd z}J_{9},\\
&[J_{1},J_{8}]=f\left(2z\frac{\rmd}{\rmd z}-1\right)J_{1}
 +\left(\frac{2f'}{f''}\frac{\rmd}{\rmd z}+1\right)J_{9},
\end{align*}
\begin{align*}
&{}[J_{2},J_{3}]=\left(2\frac{z f'-f}{f''}\frac{\rmd}{\rmd z}+z
 \right)J_{1},\quad [J_{2},J_{4}]=\left(2\frac{z f'-f}{f''}
 \frac{\rmd}{\rmd z}+1\right)J_{1},\\
&{}[J_{2},J_{5}]=z\left(2\frac{zf''-f'}{f''}\frac{\rmd}{\rmd z}+1\right)
 J_{1}+\frac{2z}{f''}\frac{\rmd}{\rmd z}J_{4},\\
&[J_{2},J_{6}]=f\left(2\frac{zf''-f'}{f''}\frac{\rmd}{\rmd z}+1\right)
 J_{1}+z\left(\frac{2f'}{f''}\frac{\rmd}{\rmd z}+1\right)J_{4},\\
&[J_{2},J_{7}]=2z\frac{z^{2}f''-zf'+f}{f''}\frac{\rmd}{\rmd z}J_{1}
 +\frac{2z}{f''}\frac{\rmd}{\rmd z}J_{9},\\
&[J_{2},J_{8}]=2f\frac{z^{2}f''-zf'+f}{f''}\frac{\rmd}{\rmd z}J_{1}
 +z\left(\frac{2f'}{f''}\frac{\rmd}{\rmd z}+1\right)J_{9},
\end{align*}
\begin{align*}
&[J_{3},J_{4}]=2\frac{ff''-(f')^{2}}{f''}\frac{\rmd}{\rmd z}J_{1},\quad
 [J_{3},J_{5}]=2z\frac{ff''-(f')^{2}}{f''}\frac{\rmd}{\rmd z}J_{1}
 +\frac{2f}{f''}\frac{\rmd}{\rmd z}J_{4},\\
&[J_{3},J_{6}]=2f\frac{ff''-(f')^{2}}{f''}\frac{\rmd}{\rmd z}J_{1}
 +f\left(\frac{2f'}{f''}\frac{\rmd}{\rmd z}+1\right)J_{4},\\
&[J_{3},J_{7}]=2z\frac{zff''-z(f')^{2}+ff'}{f''}\frac{\rmd}{\rmd z}J_{1}
 +\frac{2f}{f''}\frac{\rmd}{\rmd z}J_{9},\\
&[J_{3},J_{8}]=2f\frac{zff''-z(f')^{2}+ff'}{f''}\frac{\rmd}{\rmd z}J_{1}
 +f\left(\frac{2f'}{f''}\frac{\rmd}{\rmd z}+1\right)J_{9},
\end{align*}
\begin{align*}
&[J_{4},J_{5}]=\left(\frac{2f'}{f''}\frac{\rmd}{\rmd z}-1\right)J_{4},\quad
 [J_{4},J_{6}]=\frac{2(f')^{2}}{f''}\frac{\rmd}{\rmd z}J_{4},\\
&[J_{4},J_{7}]=2zf\frac{\rmd}{\rmd z}J_{1}-z J_{4}+\left(\frac{2f'}{f''}
 \frac{\rmd}{\rmd z}-1\right)J_{9},\\
&[J_{4},J_{8}]=2f^{2}\frac{\rmd}{\rmd z}J_{1}-f J_{4}+\frac{2(f')^{2}}{f''}
 \frac{\rmd}{\rmd z}J_{9},
\end{align*}
\begin{align*}
&[J_{5},J_{6}]=\left(2f'\frac{zf'-f}{f''}\frac{\rmd}{\rmd z}+f\right)J_{4},\\
&[J_{5},J_{7}]=2z^{2}f\frac{\rmd}{\rmd z}J_{1}-2z\frac{zf'-f}{f''}
 \frac{\rmd}{\rmd z}J_{4}+z\left(\frac{2f'}{f''}\frac{\rmd}{\rmd z}-1
 \right)J_{9},\\
&[J_{5},J_{8}]=2zf^{2}\frac{\rmd}{\rmd z}J_{1}-2f\frac{zf'-f}{f''}
 \frac{\rmd}{\rmd z}J_{4}+\frac{2z(f')^{2}}{f''}\frac{\rmd}{\rmd z}J_{9},
\end{align*}
\begin{align*}
&[J_{6},J_{7}]=2zf^{2}\frac{\rmd}{\rmd z}J_{1}-2zf'\frac{zf'-f}{f''}
 \frac{\rmd}{\rmd z}J_{4}+f\left(\frac{2f'}{f''}\frac{\rmd}{\rmd z}-1
 \right)J_{9},\\
&[J_{6},J_{8}]=2f^{3}\frac{\rmd}{\rmd z}J_{1}-2ff'\frac{zf'-f}{f''}
 \frac{\rmd}{\rmd z}J_{4}+\frac{2f(f')^{2}}{f''}\frac{\rmd}{\rmd z}J_{9},\\
&[J_{7},J_{8}]=2\frac{z^{2}(f')^{2}-2zff'+f^{2}}{f''}\frac{\rmd}{\rmd z}
 J_{9}.
\end{align*}

\section{The Explicit Forms of the Coefficients}
\label{app:coef}

The constant coefficients $C_{ij}(\alpha)$ ($i=1,\dots,4;\,j=0,\dots,8$)
appeared in (\ref{eq:JsX2}) and (\ref{eq:KsX2}) are explicitly given by
\begin{align}
\begin{split}
&C_{12}(\alpha)=2(\alpha+3),\quad C_{13}(\alpha)=-2,\quad
 C_{14}(\alpha)=-(\alpha+2),\\
&C_{15}(\alpha)=1,\quad C_{10}(\alpha)=-2,
\end{split}
\end{align}
\begin{align}
\begin{split}
&C_{21}(\alpha)=\frac{2(\alpha+3)(\alpha+2)(\alpha-1)}{\alpha+1},\quad
 C_{22}(\alpha)=-\frac{2(\alpha-1)(3\alpha^{2}+12\alpha+13)}{
 \alpha+1},\\
&C_{23}(\alpha)=\frac{2(\alpha^{2}-3\alpha-2)}{
 \alpha(\alpha+1)},\quad C_{24}(\alpha)=\frac{(\alpha+2)(\alpha-1)
 (3\alpha^{2}+6\alpha+4)}{\alpha(\alpha+1)},\\
&C_{25}(\alpha)=\frac{4(\alpha+2)}{\alpha(\alpha+1)},\quad
 C_{26}(\alpha)=-\frac{\alpha}{\alpha+1},\quad C_{27}(\alpha)=\frac{
 2(\alpha-1)}{\alpha},\\
&C_{20}(\alpha)=\frac{2(\alpha-2)(\alpha-1)}{\alpha},
\end{split}
\end{align}
\begin{align}
\begin{split} 
&C_{33}(\alpha)=2(3\alpha^{2}+7\alpha+6),\quad C_{35}(\alpha)=-(
 3\alpha^{2}+5\alpha+4),\quad C_{36}(\alpha)=\alpha,\\
&C_{37}(\alpha)=-2(\alpha-1),\quad C_{30}(\alpha)=4(\alpha+4)(\alpha-1),
\end{split}
\end{align}
\begin{align}
\begin{split}
&C_{42}(\alpha)=-2\alpha(\alpha+3)^{2}(\alpha-1),\quad C_{43}(\alpha)=
 -\frac{(\alpha-1)(7\alpha^{3}+31\alpha^{2}+54\alpha+36)}{\alpha+1},\\
&C_{44}(\alpha)=\frac{\alpha(\alpha+3)(\alpha+2)^{2}(\alpha-1)}{\alpha+1},
 \quad C_{45}(\alpha)=\frac{(\alpha-1)(7\alpha^{3}+31\alpha^{2}+54\alpha+48)
 }{2(\alpha+1)},\\
&C_{46}(\alpha)=-\frac{(3\alpha^{3}-5\alpha^{2}
 -14\alpha-8)}{2(\alpha+1)},\quad C_{47}(\alpha)=\frac{(\alpha-1)
 (3\alpha^{2}+\alpha-12)}{\alpha+1},\\
&C_{48}(\alpha)=\frac{2(\alpha-1)}{\alpha+1},\quad C_{40}(\alpha)=-\frac{
 4(\alpha-1)(\alpha^{3}+7\alpha^{2}-10)}{\alpha+1}.
\end{split}
\end{align}
The other coefficients which are not appeared above all vanish.


\bibliography{refsels}
\bibliographystyle{npb}



\end{document}